\documentclass[oneside,showpacs,final,onecolumn,10pt,titlepage,pra,preprint]{revtex4}
\usepackage{amsfonts}
\usepackage{amsmath}
\usepackage{amssymb}
\usepackage{graphicx}
\usepackage{dcolumn}
\usepackage{bm}

\input{tcilatex}

\begin{document}

\preprint{}
\title{Relaxation and decoherence in a resonantly driven qubit}
\author{Zhongyuan Zhou$^{1,2}$, Shih-I Chu$^{1}$, and Siyuan Han$^{2}$}
\affiliation{$^{1}$Department of Chemistry, University of Kansas, Lawrence, KS 66045\\
$^{2}$Department of Physics and Astronomy, University of Kansas, Lawrence,
KS 66045}

\begin{abstract}
Relaxation and decoherence of a qubit coupled to environment and driven by a
resonant ac field are investigated by analytically solving Bloch equation of
the qubit. It is found that the decoherence of a driven qubit can be
decomposed into intrinsic and field-dependent ones. The intrinsic
decoherence time equals to the decoherence time of the qubit in free decay
while the field-dependent decoherence time is identical with the relaxation
time of the qubit in driven oscillation. Analytical expressions of the
relaxation and decoherence times are derived and applied to study a
microwave-driven SQUID flux qubit. The results are in excellent agreement
with those obtained by numerically solving the master equation. The
relations between the relaxation and decoherence times of a qubit in free
decay and driven oscillation can be used to extract the decoherence and thus
dephasing times of the qubit by measuring its population evolution in free
decay and resonantly driven oscillation.
\end{abstract}

\maketitle

\section{Introduction}

In recent years, significant progress has been made on physical
implementation of quantum computation based on superconducting qubits.
Quantum coherence has been successfully demonstrated in a variety of
superconducting single-qubit systems \cite%
{Nakamura1999,Vion2002,Yu2002,Martinis2002,Friedman2000,Wal2000,Chiorescu2003,Chiorescu2004}
and coupled two-qubit systems, \cite%
{Pashkin2003,Yamamoto2003,Berkley2003,McDermott05} indicating the potential
of superconducting qubits in quantum computing. However, due to unavoidable
coupling with environment, the superconducting (charge, flux, and phase)
qubits always suffer from decoherence such as relaxation and dephasing,
resulting in relatively short coherence times. For this reason,
environment-induced decoherence has been and is still a main obstacle to the
practical application of superconducting qubits in quantum computation. \cite%
{Vion2002,Astafiev04,Ithier05,Makhlin2001,Mooij1999}

The environment-induced decoherence of superconducting qubits has been
extensively studied both theoretically\emph{\ }\cite%
{Mooij1999,Makhlin2001,Burkard05-1,Goorden04,Makhlin04,Orlando2002,Harlingen04,Zhou2004,Shresta05,Anastopoulos2000,Robertson05,Cheng2004,Storcz2003,Thorwart2001,Governale2001,Tian2002,Xu2005,Hartmann2000,Falci05}
and experimentally\emph{\ }\cite%
{Chiorescu2004,Astafiev04,Nakamura02,Lehnert03,Duty04,Lishaoxiong06,Bertet06,Vion2002,Berkley03-1,Dutta2004,Harlingen04,Robertson05}
in the absence of ac driving fields (free decay). Quite a few proposals,
such as dynamical decoupling, \cite%
{Viola05,Viola98,Viola99,Vitali02,Gutmann05,Falci04,Faoro04,Shiokawa04}
decoherence free subspaces, \cite{Duan97,Zanardi97,Lidar98,Beige00} spin
echoes, \cite{Chiorescu2003,Vion2002,Nakamura02} and coherence-preserving
qubits, \cite{Bacon01} have been proposed to reduce such kind of
decoherence. However, in superconducting-qubit based quantum computation, ac
fields (e.g., microwave fields) are usually used to manipulate the qubit's
state. \cite%
{Nakamura1999,Vion2002,Yu2002,Martinis2002,Friedman2000,Wal2000,Chiorescu2003,Pashkin2003,Yamamoto2003,Berkley2003,Chiorescu2004,McDermott05,Goorden04}
Recent experiment \cite{Ithier05} shows that the decoherence time of a
superconducting qubit is significantly increased in the presence of a
resonantly ac driving field. Thus a comprehensive understanding to
decoherence of a realistic superconducting qubit needs to include influence
of driving fields. \cite{Goorden04}

In this paper, we study the effect of driving fields on the relaxation and
decoherence of a driven qubit. We focus our study on weak resonant driving
fields characterized by $\Omega \ll \omega _{\mu }$ (where, $\Omega $ and $%
\omega _{\mu }$ are the Rabi frequency and the frequency of ac driving
field, respectively) and low temperatures. In this case, population leakage
to non-computational states due to strong field effect \cite%
{zhou2005PRL-ITA,zhou2006} and thermal activation \cite{Burkard2004} is
negligible and thus a multilevel superconducting qubit coupled to
environment and driven by a resonant ac field can be well approximated by a
resonantly driven dissipative \emph{two-level system} (TLS). We first
explore relaxation and decoherence of a qubit in free decay and demonstrate
that our results agree very well with those in the literature. We then
derive analytical expressions of relaxation and decoherence times for a
driven qubit through analytical solutions of Bloch equation of the driven
dissipative TLS. The relations between the relaxation and decoherence times
can be used to extract decoherence and dephasing times of the qubit by
measuring its population evolution in free decay and driven oscillation.
Finally, we use the analytical expressions to study relaxation and
decoherence of a superconducting quantum interface device (SQUID) flux qubit
driven by a microwave and show that the analytical results are in excellent
agreement with the results obtained by numerically solving the master
equation.

\section{Bloch equation of a general driven dissipative two-level system}

A dissipative system can be described by a reduced density operator. In the
Hilbert space spanned by the eigenstate $\left\vert m\right\rangle $ of
eigenenergy $E_{m}$ $(m=1,2,\cdots )$, the reduced density operator $%
\widehat{\rho }$ is represented by a reduced density matrix with matrix
elements $\rho _{mn}=\left\langle m\left\vert \widehat{\rho }\right\vert
n\right\rangle $. The diagonal matrix element $\rho _{mm}$ and off-diagonal
matrix element $\rho _{mn}$ $(m\neq n)$ are the population and coherence of
the system, respectively.

In general, the reduced density operator is governed by generalized master
equation of non-Markovian process. \cite%
{Brinati94,Smirnov2003,Hartmann2000,Shresta05} However, in the case of weak
damping and low temperature considered here, the generalized non-Markovian
master equation is equivalent to the Markovian master equation \cite%
{Hartmann2000} and in the case of resonant driving, the results obtained
from the Markovian master equation are the same as those from the
non-Markovian master equation. \cite{Brinati94} Furthermore, in the case of
weak damping, Lamb shifts are usually very small compared to the ac driving
field and thus are often neglected \cite{Louisell1973}. After dropping the
terms related to the Lamb shifts, the Markovian master equation for the
reduced density matrix of a driven multilevel qubit is cast into the
generalized Bloch-Redfield equation. \cite{Weiss1999,Burkard2004,Zhou-PRB-I}
For a driven dissipative TLS with conserved population, the reduced density
matrix elements satisfy $\rho _{21}=\rho _{12}^{\ast }$ and $\rho _{11}+\rho
_{22}=1$ and are governed by \cite{Burkard2004,Zhou-PRB-I}%
\begin{eqnarray}
\frac{d\rho _{11}}{dt} &=&-iH_{12}^{F}\left( \rho _{21}-\rho _{12}\right)
-R_{22,11}\rho _{11}  \notag \\
&&+R_{11,12}\left( \rho _{12}+\rho _{21}\right) +R_{11,22}\rho _{22},
\label{1} \\
\frac{d\rho _{12}}{dt} &=&i\omega _{21}\rho _{12}-i\left[ H_{12}^{F}\left(
\rho _{22}-\rho _{11}\right) \right.  \notag \\
&&+\left. \left( H_{11}^{F}-H_{22}^{F}\right) \rho _{12}\right]
+R_{12,11}\rho _{11}  \notag \\
&&+R_{12,12}\rho _{12}+R_{12,21}\rho _{21}+R_{12,22}\rho _{22},  \label{2}
\end{eqnarray}%
where, $\omega _{mn}=\left( E_{m}-E_{n}\right) /\hbar $\ is the transition
frequency, $R_{mn,m^{\prime }n^{\prime }}$ is the matrix element of damping
rate superoperator describing the effect of environment on the TLS, \cite%
{Burkard2004,Zhou-PRB-I} and $H_{mn}^{F}=\left\langle m\left\vert
H_{F}\right\vert n\right\rangle /\hbar $ is the matrix element of
interaction Hamiltonian $H_{F}$ between the TLS and resonant ac driving
field. For simplicity and without loss of generality, henceforward, unless
otherwise specified, we assume $H_{12}^{F}=\varepsilon \cos \left( \omega
_{\mu }t\right) $ and $H_{11}^{F}-H_{22}^{F}=\delta \cos \left( \omega _{\mu
}t\right) $, where $\varepsilon $ and $\delta $ are the constants associated
with transition matrix elements and field strength.

The damping of a dissipative system is caused by the interaction between the
system and environment such as a thermal bath. In the case of weak damping,
the interaction Hamiltonian $H_{I}$ between the system and the thermal bath
can be approximated by a linear function of system variable and expressed as 
\cite{Weiss1999,Makhlin2001,Xu2005} $H_{I}=-\Lambda x\mathcal{U}_{B}$,
where, $x$ is the generalized coordinate of system, $\mathcal{U}_{B}$ is a
function of bath variable(s), and $\Lambda $ is a coupling constant. The
effect of the thermal bath on the system can equivalently be characterized
by a spectral density $J\left( \omega \right) $. \cite{Leggett87,Devoret97}
In this case, the matrix elements of the damping rate superoperator can be
calculated by \cite{Burkard2004,Zhou-PRB-I}%
\begin{eqnarray}
R_{mn,m^{\prime }n^{\prime }} &=&\frac{1}{2\hbar ^{2}\Lambda ^{2}}\left[
-\delta _{nn^{\prime }}\sum_{k}x_{mk}x_{km^{\prime }}J\left( \omega
_{m^{\prime }k}\right) \right.  \notag \\
&&+x_{mm^{\prime }}x_{n^{\prime }n}\left[ J\left( \omega _{n^{\prime
}n}\right) +J\left( \omega _{m^{\prime }m}\right) \right]  \notag \\
&&-\left. \delta _{mm^{\prime }}\sum_{k}x_{n^{\prime }k}x_{kn}J\left( \omega
_{n^{\prime }k}\right) \right] ,  \label{i-0}
\end{eqnarray}%
where, $x_{mn}$ is the transition matrix element.

An equivalent approach to describe the TLS is to use Bloch vector $\left(
u,v,w\right) $ defined by%
\begin{eqnarray}
u &=&Tr\left( \rho \sigma _{x}\right) =2\text{Re}\rho _{12},  \label{i-1} \\
v &=&Tr\left( \rho \sigma _{y}\right) =-2\text{Im}\rho _{12},  \label{i-2} \\
w &=&Tr\left( \rho \sigma _{z}\right) =\rho _{11}-\rho _{22}.  \label{i-3}
\end{eqnarray}%
The components of the Bloch vector $u$ and $v$ represent the real and
imaginary parts of the coherence $\rho _{12}$, while the component $w$ is
the population difference. Substituting the Bloch vector into the master
equation (\ref{1}) and (\ref{2}) one obtains the following Bloch equation%
\begin{eqnarray}
\frac{du}{dt} &=&\left( R_{12,12}+R_{12,21}\right) u+\left[ \omega
_{21}-\delta \cos \left( \omega _{\mu }t\right) \right] v  \notag \\
&&+\left( R_{12,11}-R_{12,22}\right) w+\left( R_{12,11}+R_{12,22}\right) ,
\label{m-2} \\
\frac{dv}{dt} &=&-\left[ \omega _{21}-\delta \cos \left( \omega _{\mu
}t\right) \right] u+\left( R_{12,12}-R_{12,21}\right) v  \notag \\
&&-2\varepsilon \cos \left( \omega _{\mu }t\right) w,  \label{m-2-1} \\
\frac{dw}{dt} &=&2R_{11,12}u+2\varepsilon \cos \left( \omega _{\mu }t\right)
v-\left( R_{22,11}+R_{11,22}\right) w  \notag \\
&&+\left( R_{11,22}-R_{22,11}\right) .  \label{m-2-2}
\end{eqnarray}

\section{Relaxation and decoherence in the absence of ac driving fields}

Suppose the initial value of the Bloch vector is $\left(
u_{0},v_{0},w_{0}\right) $. In the absence of ac driving fields (free decay)
and under the rotating-wave approximation (RWA), the solution of the Bloch
equation is given by (see Appendix A for details)%
\begin{eqnarray}
u &=&\left[ A\cos \left( \varpi t-\theta _{0}\right) +B\sin \left( \varpi
t\right) \right] e^{-\kappa t},  \label{m30-3} \\
v &=&\left[ -A\sin \left( \varpi t\right) +B\cos \left( \varpi t+\theta
_{0}\right) \right] e^{-\kappa t},  \label{m30-4} \\
w &=&w_{\infty }+De^{-\gamma t},  \label{m-30}
\end{eqnarray}%
where, $\kappa =-R_{12,12}$ is introduced for convenience, $\gamma
=R_{22,11}+R_{11,22}$ is the relaxation rate in free decay, $w_{\infty }$, $%
\varpi $, and $\theta _{0}$ are the parameters related to the damping rate
matrix elements, and $A$, $B$, and $D$ are the constants determined by the
initial condition of the TLS and damping rate matrix elements. These
parameters and constants are given by Eqs. (\ref{i-4-4}) to (\ref{i-4-3}) in
Appendix A. It is shown that the population and coherence undergo simple
exponential decays independently at rates $\gamma $ and $\kappa $,
respectively, and the coherence components $u$ and $v$ contain fast
oscillating factors of frequency $\varpi $. When $t\rightarrow \infty $ the
coherence components tend to zero, while the population difference tends to
a constant $w_{\infty }$. In the case of zero damping ($R_{12,21}=0$) $%
\varpi =\omega _{21}$ and $\theta _{0}=0$ while in the case of weak damping (%
$R_{12,21}\ll \omega _{21}$) $\varpi <\omega _{21}$ and $\theta _{0}>0$.
Hence $w_{\infty }$ is the population difference at $t\rightarrow \infty $, $%
\varpi $ is the transition frequency modified by the damping, and $\theta
_{0}$ is the phase shift due to the damping.

It is shown from Eq. (\ref{m-30}) that the population (difference) decays
exponentially with the rate of $\gamma $. Hence, the relaxation time is%
\begin{equation}
T_{1}=\gamma ^{-1}.  \label{x-80}
\end{equation}%
From Eqs. (\ref{m30-3}) and (\ref{m30-4}) the coherence, $\rho
_{12}=(u-iv)/2\propto e^{-\kappa t}$, decays exponentially with the rate of $%
\kappa $. The decoherence time is therefore 
\begin{equation}
T_{2}=\kappa ^{-1}.  \label{x-81}
\end{equation}%
These results agree very well with those obtained by others. \cite%
{Burkard2004} Eqs. (\ref{x-80}) and (\ref{x-81}) indicate that for free
decay $T_{1}$ is independent of $T_{2}$. In general $T_{2}<2T_{1}$ due to
dephasing. \cite{Falci05} The pure dephasing time $T_{\varphi }$ is related
to $T_{1}$ and $T_{2}$ by \cite{Burkard2004}%
\begin{equation}
\frac{1}{T_{\varphi }}=\frac{1}{T_{2}}-\frac{1}{2T_{1}}.  \label{x-81-1}
\end{equation}

For the general case of a qubit coupled to a thermal bath, using Eq. (\ref%
{i-0}), we obtain%
\begin{eqnarray}
T_{1}^{-1} &=&\frac{\left\vert x_{12}\right\vert ^{2}}{\hbar ^{2}\Lambda ^{2}%
}\left[ J\left( \omega _{21}\right) +J\left( \omega _{12}\right) \right] ,
\label{x-81-1-0} \\
T_{2}^{-1} &=&\frac{1}{2T_{1}}+\frac{\left( x_{11}-x_{22}\right) ^{2}}{%
2\hbar ^{2}\Lambda ^{2}}J\left( 0\right) ,  \label{x-81-1-1}
\end{eqnarray}%
and%
\begin{equation}
T_{\varphi }^{-1}=\frac{\left( x_{11}-x_{22}\right) ^{2}}{2\hbar ^{2}\Lambda
^{2}}J\left( 0\right) .  \label{x-81-1-2}
\end{equation}%
These results demonstrate that the relaxation and dephasing rates of the a
qubit are determined by the spectral densities at the transition frequency $%
\omega =\left\vert \omega _{21}\right\vert $ and low frequency $\omega =0$,
respectively. \cite%
{Burkard2004,Makhlin04,Ithier05,Bertet06,Berkley03-1,Xu2005} In addition,
the relaxation rate is proportional to the modulus square of the transition
matrix element $\left\vert x_{12}\right\vert ^{2}$, while the dephasing rate
is proportional to the square of the difference of average coordinates of
the two states $\left( x_{11}-x_{22}\right) ^{2}$. For a qubit having $%
\left( x_{11}-x_{22}\right) =0$, the dephasing is completely suppressed.

\section{Relaxation and decoherence in the presence of an ac driving field}

For a driven dissipative TLS, $\varepsilon \neq 0$ and also $\delta \neq 0$
in general. Under the RWA, the solution of the Bloch equation in the
underdamped regime with $\varepsilon >\eta $ is given by (see Appendix B for
details)%
\begin{eqnarray}
u &=&A\cos \left( \varpi t-\theta _{0}\right) e^{-\kappa t}+B_{0}\sin \left(
\varpi t\right)  \notag \\
&&+B_{1}\sin \left( \Omega t+\theta _{1}\right) \sin \left( \varpi t\right)
e^{-\Gamma t},  \label{x-75} \\
v &=&-A\sin \left( \varpi t\right) e^{-\kappa t}+B_{0}\allowbreak \cos
\left( \varpi t+\theta _{0}\right)  \notag \\
&&+B_{1}\sin \left( \Omega t+\theta _{1}\right) \cos \left( \varpi t+\theta
_{0}\right) e^{-\Gamma t},  \label{x-76} \\
w &=&D_{0}+D_{1}\sin \left( \Omega t+\theta _{2}\right) \allowbreak
e^{-\Gamma t},  \label{x-77}
\end{eqnarray}%
where, $\eta =\left( \gamma -\kappa \right) /2$ is the damping strength, $%
\Omega =\sqrt{\varepsilon ^{2}-\eta ^{2}}$ is the Rabi frequency, $\Gamma
=\left( \gamma +\kappa \right) /2$ is the mean value of relaxation and
decoherence rates in free decay, $A$ is the field-independent constant
identical with that in Eqs. (\ref{m30-3}) and (\ref{m30-4}) and is given by
Eq. (\ref{i-0-1}) in Appendix A, $B_{0}$, $B_{1}$, $D_{0}$, $D_{1}$, $\theta
_{1}$, and $\theta _{2}$ are the field-dependent constants given by Eqs. (%
\ref{a-5}) to (\ref{x-7}) in Appendix B. For quantum information processing,
manipulation of qubits must be in the underdamped regime. Hence, hereafter
our discussion is focused on the resonantly driven qubit in this regime.

It is shown clearly that the solution given by Eqs. (\ref{x-75}) to (\ref%
{x-77}) contains a slow Rabi oscillation. It is also shown that in general
the population and coherence undergo more complicated damped oscillations
which are totally different from those in free decay. The first terms of the
coherence components on the right-hand sides of Eqs. (\ref{x-75}) and (\ref%
{x-76}) are independent of the driving field. They only depend on the
initial state and matrix elements of the damping rate superoperator. Hence
they represent intrinsic decoherence induced by the coupling between the
qubit and thermal bath. Note that the intrinsic decoherence rate $\kappa $
of a qubit in driven oscillation equals to the decoherence rate of free
decay (see Eqs. (\ref{m30-3}) and (\ref{m30-4})). In contrast, the remaining
terms of the coherence components on the right-hand sides of Eqs. (\ref{x-75}%
) and (\ref{x-76}) strongly depend on the driving field. In particular, the
third terms decay exponentially and thus represent field-dependent
decoherence. The field-dependent decoherence rate $\Gamma $ and the
relaxation rate of the qubit in driven oscillation are equal (see Eq. (\ref%
{x-77})).

Note that both the relaxation and decoherence should have been eliminated in
principle if $A=B_{1}=D_{1}=0$ in Eqs. (\ref{x-75}) to (\ref{x-77}).
However, such conditions can not be fulfilled in reality for a resonantly
driven qubit starting from a pure state. Thus the relaxation and decoherence
can not be suppressed simultaneously with a single resonant driving field. 
\cite{Viola98,Viola99,Vitali02,Gutmann05,Falci04,Faoro04} Nevertheless, as
will be demonstrated below, for a special class of initial states, a
properly chosen resonant driving field could slow down the decoherence. \cite%
{Grifoni1999}

Since in general the population and coherence of a driven qubit decay with
more than one rates, it is necessary to introduce multiple relaxation and
decoherence times to describe the system's behavior properly. Each
relaxation time or decoherence time represents one of the characteristic
times of the qubit for each specific case. For instance, from Eq. (\ref{x-77}%
), the relaxation time $\widetilde{T}_{1}$ is defined by%
\begin{equation}
\widetilde{T}_{1}=\Gamma ^{-1},  \label{x-82}
\end{equation}%
where the tilde "$\sim $" is used to indicate that the qubit is resonantly
driven. Similarly, from\ Eqs. (\ref{x-75}) and (\ref{x-76}), the intrinsic
decoherence time $\widetilde{T}_{2,1}$ and field-dependent decoherence time $%
\widetilde{T}_{2,2}$ are defined by%
\begin{equation}
\widetilde{T}_{2,1}=\kappa ^{-1}=T_{2},  \label{x-84}
\end{equation}%
and%
\begin{equation}
\widetilde{T}_{2,2}=\Gamma ^{-1}=\widetilde{T}_{1},  \label{x-84-1}
\end{equation}%
respectively. Namely, the intrinsic decoherence time $\widetilde{T}_{2,1}$
and the field-dependent decoherence time $\widetilde{T}_{2,2}$ of a qubit in
driven oscillation equal to the decoherence time $T_{2}$ of the qubit in
free decay and the relaxation time $\widetilde{T}_{1}$ of the qubit in
driven oscillation, respectively. If the initial state of the driven qubit
corresponds to $u_{0}=0$ (e.g., the ground state), then the intrinsic
decoherence terms vanish and the decoherence of the qubit is characterized
completely by the field-dependent decoherence time $\widetilde{T}_{2,2}$.
Note that the Eqs. (\ref{x-81}) and (\ref{x-82}) are used to obtain the
second equality of Eqs. (\ref{x-84}) and (\ref{x-84-1}).

From Eqs. (\ref{x-80}) to (\ref{x-81-1}), Eq. (\ref{x-84}), and Eq. (\ref%
{x-84-1}) we obtain%
\begin{equation}
\frac{1}{\widetilde{T}_{2,2}}=\frac{1}{\widetilde{T}_{1}}=\frac{1}{2T_{1}}+%
\frac{1}{2T_{2}}=\frac{3}{4T_{1}}+\frac{1}{2T_{\varphi }}.  \label{x-86}
\end{equation}%
It indicates that $\widetilde{T}_{2,2}=\widetilde{T}_{1}$ is always in
between $T_{1}$ and $T_{2}$ and smaller than the lesser of $4T_{1}/3$ and $%
2T_{\varphi }$. This prediction is similar to those obtained by others \cite%
{Shresta05,Anastopoulos2000,Kosugi05} and also agrees, within the
experimental uncertainties, with the results of recent experiment. \cite%
{Ithier05}

The relations between various characteristic times are summarized in TABLE %
\ref{Table 1}. It is clearly shown that for a driven qubit the intrinsic
decoherence time is independent of the driving field and is identical with
the decoherence time of the qubit in free decay. In contrast, the relaxation
and field-dependent decoherence times of the driven qubit strongly depend on
the driving field and are equal to each other.

\begin{table}[htbp] \centering%
\caption{Relaxation and decoherence times of a qubit in free and driven decays.\label{Table
1}}%
\begin{tabular}{ccccc}
\hline\hline
Decay type & \ \ \  & Relaxation & \ \ \ \  & Decoherence \\ \cline{5-5}
\  & \  & \  & \  & 
\begin{tabular}{lll}
\ intrinsic \  & \ \  & field-dependent%
\end{tabular}
\\ \hline
\begin{tabular}{c}
Free decay \\ 
Driven decay%
\end{tabular}
& 
\begin{tabular}{l}
\  \\ 
\ \ 
\end{tabular}
& 
\begin{tabular}{l}
$T_{1}=\gamma ^{-1}$ \\ 
$\widetilde{T}_{1}=\Gamma ^{-1}$%
\end{tabular}
& 
\begin{tabular}{l}
\  \\ 
\ \ 
\end{tabular}
& 
\begin{tabular}{lll}
$T_{2}=\kappa ^{-1}$ \ \ \ \  &  &  \\ 
$\widetilde{T}_{2,1}=T_{2}$ &  & $\widetilde{T}_{2,2}=\widetilde{T}_{1}$%
\end{tabular}
\\ \hline\hline
\end{tabular}%
\end{table}%

The relations between the relaxation and decoherence times of a qubit in
free decay and driven oscillation can be used to extract the decoherence and
dephasing times of the qubit by measuring its population evolution. In
experiment, one would first measure the population evolution of the qubit in
the free decay and in resonantly driven oscillation (Rabi oscillation) to
obtain $T_{1}$ and $\widetilde{T}_{1}$. Then the decoherence times $T_{2}$, $%
\widetilde{T}_{2,1}$, and $\widetilde{T}_{2,2}$ can be evaluated using Eqs. (%
\ref{x-84}) to (\ref{x-86}). Finally, the pure dephasing time of the qubit $%
T_{\varphi }$ can be calculated from Eq. (\ref{x-81-1}).

\section{Microwave-driven SQUID flux qubits: Analytical versus numerical
results}

To demonstrate validity of the analytical expressions of relaxation and
decoherence times obtained in the preceding sections we apply them to
calculate relaxation and decoherence times of a microwave-driven SQUID flux
qubit inductively coupled to environment and compare the results with those
obtained by numerically solving the master equation given by Eqs. (\ref{1})
and (\ref{2}). An rf SQUID consists of a superconducting loop of inductance $%
L$\ interrupted by a Josephson tunnel junction (JJ) which, applying the
resistively-shunted junction (RSJ) model \cite{Danilov1983}, is
characterized by the critical current $I_{c}$, shunt capacitance $C$, and
shunt resistance $R$. The SQUID flux qubit is usually coupled to its control
and readout circuits. A typical equivalent circuit is shown in FIG. 1 (a)
along with its equivalent admittance $Y(\omega )$ in FIG. 1 (b). In this
simplified external circuit the left part is the SQUID and the right part
supplies external flux to the SQUID qubit. For a superconducting device such
as a SQUID, the thermal bath is the external circuit coupled to the device.
Thus the external circuit is the dominant source of dissipation for the
SQUID qubit \cite{Makhlin2001}. In this section, we analyze the relaxation
and decoherence of the SQUID flux qubit due to coupling to the external
circuit.

\subsection{Hamiltonian of a microwave-driven SQUID flux qubit}

The Hamiltonian of a flux-biased rf SQUID with total magnetic flux $\Phi $
enclosed in the loop can be written as \cite{Zhou2002,Zhou2004} 
\begin{equation}
H_{0}\left( x\right) =\frac{p^{2}}{2m}+V\left( x\right) ,  \label{b0}
\end{equation}%
where, $m=C\Phi _{0}^{2}$ is the mass of \textquotedblleft
flux\textquotedblright\ particle, $\Phi _{0}\equiv h/2e$ is the flux
quantum, $e$ is the elementary charge, $x=\Phi /\Phi _{0}$ is the canonical
coordinate of \textquotedblleft flux\textquotedblright\ particle, $p=-i\hbar
\partial /\partial x$ is the canonical momentum conjugate to $x$, and $V(x)$
is the potential energy given by%
\begin{equation}
V\left( x\right) =\frac{1}{2}m\omega _{LC}^{2}\left( x-x_{e}\right)
^{2}-E_{J}\cos \left( 2\pi x\right) .  \label{b1}
\end{equation}%
Here, $E_{J}=\hbar I_{c}/2e=m\omega _{LC}^{2}\beta _{L}/4\pi ^{2}$ is the
Josephson coupling energy, $\beta _{L}=2\pi LI_{c}/\Phi _{0}$ is the
potential shape parameter, $\omega _{LC}=1/\sqrt{LC}$ is the characteristic
frequency of the SQUID, and $x_{e}=\Phi _{e}/\Phi _{0}$ is the normalized
external fluxes from the external circuit.

To perform gate operations, a microwave pulse is applied to the SQUID qubit.
If the interaction between the microwave and external circuit is negligible
the Hamiltonian of the microwave-driven SQUID qubit coupled to the external
circuit is given by%
\begin{equation}
H\left( x,t\right) =H_{0}\left( x\right) +H_{F}\left( x,t\right) +H_{I},
\label{b1-1}
\end{equation}%
where, $H_{F}\left( x,t\right) $ is the interaction Hamiltonian between the
SQUID qubit and microwave and $H_{I}$ is the interaction Hamiltonian between
the SQUID qubit and the external circuit (thermal bath).

If $\phi \left( t\right) $ is the normalized flux coupled to the SQUID from
the microwave then $H_{F}\left( x,t\right) $ is given by \cite{Zhou2002}%
\begin{equation}
H_{F}\left( x,t\right) =\frac{m\omega _{LC}^{2}}{2}\phi \left[ \phi +2\left(
x-x_{e}\right) \right] .  \label{b4}
\end{equation}%
Hereafter $\phi $ is taken to be%
\begin{equation}
\phi (t)=\phi _{\mu }\cos \left( \omega _{\mu }t\right) ,  \label{b4-1}
\end{equation}%
where, $\phi _{\mu }$ and $\omega _{\mu }$ are the field strength and
frequency of the microwave, respectively. From Eqs. (\ref{b4}) and (\ref%
{b4-1}) one has%
\begin{equation}
\varepsilon =\frac{m\omega _{LC}^{2}\phi _{\mu }\left\vert x_{12}\right\vert 
}{\hbar },  \label{b4-2}
\end{equation}%
and%
\begin{equation}
\delta =\frac{m\omega _{LC}^{2}\phi _{\mu }\left( x_{11}-x_{22}\right) }{%
\hbar }.  \label{b4-3}
\end{equation}%
For a SQUID qubit, $\delta \neq 0$ if the tunneling distance $\left\vert
x_{11}-x_{22}\right\vert \neq 0$.

\subsection{Spectral density of the external circuit}

As has been addressed, in the case of weak damping, the effect of thermal
bath on the superconducting system can be described by the spectral density. 
\cite{Leggett87,Devoret97} For the SQUID flux qubit considered here $\Lambda
=-1/\Phi _{0}$ and the spectral density $J\left( \omega \right) $ is given
by \cite{Devoret97,Zhou-PRB-I}%
\begin{equation}
J\left( \omega \right) =\hbar \omega Y_{R}(\omega )\left[ 1+\coth \left( 
\frac{\hbar \omega }{2k_{B}T}\right) \right] ,  \label{i-0-0}
\end{equation}%
where, $k_{B}$ is the Boltzmann constant, $T$ is the temperature of the
thermal bath, and $Y_{R}(\omega )$ is the real part of the
frequency-dependent admittance $Y(\omega )$ of the external circuit. \cite%
{Devoret97} For the external circuit of FIG. 1 one has (see Appendix C for
details) 
\begin{equation}
Y_{R}(\omega )=\frac{F_{0}\left( \omega \right) }{\omega ^{2}+G_{0}\left(
\omega \right) },  \label{3-6}
\end{equation}%
where, $F_{0}\left( \omega \right) $ and $G_{0}\left( \omega \right) $ are
given by Eqs. (\ref{A-a10}) to (\ref{A-a14}) in Appendix C. For the sake of
concreteness, henceforth, unless otherwise specified, the SQUID flux qubit
has the parameters of $L=1.0$ nH, $C=15$ fF, $L_{e}=1.0$ nH, $C_{e}=10$ pF, $%
R_{e}=10$ $\Omega $, $R_{0}=1.0$ k$\Omega $, $\beta _{L}=1.4$, $x_{e}=0.499$%
, and $M=3.0$ pH. The temperature of the external circuit is taken to be $%
T=0.1$ K.

In FIG. 2 we plot the spectral density $J\left( \omega \right) $ versus
frequency $\omega $ for the SQUID qubit with the above parameters at $T=0.1$
K. It is shown that the spectral density $J\left( \omega \right) $ reaches
the maximum at $\omega \simeq 0.04\omega _{LC}$ and approaches a constant
value at low frequency, in particular $J\left( 0\right) \neq 0$. From Eq. (%
\ref{x-81-1-2}) $T_{\varphi }$ is finite and thus the external circuit will
induce dissipation as well as dephasing.

In FIG. 3, we show the spectral density $J\left( \omega \right) $ at $\omega
=\omega _{LC}$ versus temperature (the solid curve) together with the
quantum spectral density at $T=0$ (the dashed line) and the classical
spectral density at high temperature limit $T\gg \hbar \omega _{LC}/k_{B}$
(the dashed dotted line). The quantum spectral density is independent of
temperature while the classical spectral density is a linear function of
temperature, which agree with those in the literature. \cite{Weiss1999} Note
that for $T>1.5$ K the spectral density is approximated well by the
classical spectral density, while $T<0.3$ K the spectral density approaches
the quantum spectral density.

For the SQUID flux qubit, $Y_{R}\left( \omega \right) $ is an even function
of $\omega $, which is also the case for most of superconducting systems.
Substituting Eq. (\ref{i-0-0}) into Eqs. (\ref{x-81-1-0}) and (\ref{x-81-1-2}%
), we obtain%
\begin{equation}
T_{1}^{-1}=\frac{2\pi ^{2}}{e^{2}}\hbar \omega _{21}\left\vert
x_{12}\right\vert ^{2}Y_{R}\left( \omega _{21}\right) \coth \left( \frac{%
\hbar \omega _{21}}{2k_{B}T}\right) ,  \label{x-81-1-3}
\end{equation}%
and%
\begin{equation}
T_{\varphi }^{-1}=\frac{\pi ^{2}}{e^{2}}k_{B}T\left( x_{11}-x_{22}\right)
^{2}Y_{R}\left( 0\right) .  \label{x-81-1-4}
\end{equation}%
Thus the relaxation rate is dominated by the circuit's admittance at the
transition frequency $\omega _{21}$ while the dephasing rate by the
admittance at $\omega =0$. Furthermore, the dephasing rate is proportional
to the temperature of thermal bath. Hence at the low temperature the
dominating source of decoherence is relaxation while at the high temperature
the main source of decoherence is dephasing. These results agree with those
obtained by others. \cite{Leggett87,Tian2002,Burkard2004}

\subsection{Relaxation and decoherence times}

To numerically calculate the relaxation and decoherence times of the
microwave-driven SQUID qubit, we need to compute evolution of population and
coherence of the qubit. For this purpose we first calculate the eigenenergy $%
E_{m}$ and eigenstate $\left\vert m\right\rangle $ of the SQUID qubit by
numerically solving the time-independent Schr\"{o}dinger equation with $%
H_{0}\left( x\right) $. \cite{Zhou2002,Zhou2004} We then calculate the
transition frequency $\omega _{mn}$, damping rate matrix element $%
R_{mn,m^{\prime }n^{\prime }}$, and matrix element of interaction
Hamiltonian $H_{mn}^{F}$ between the SQUID qubit and microwave. Finally we
calculate the population and coherence by numerically solving the master
equation given by Eqs. (\ref{1}) and (\ref{2}) using the split-operator
method \cite{Hermann1988} for the algebra equation with non-symmetric
matrix. \cite{Zhou-PRB-I}

\subsubsection{Free decay of the SQUID qubit}

In general, a SQUID qubit is a multilevel system. \cite{Zhou2002} However,
when it is driven by a weak resonant microwave field, the leakage to
non-computational states is negligible \cite{zhou2005PRL-ITA,zhou2006} and
the SQUID qubit can be very well approximated by a TLS consisting of the
lowest two levels $\left\vert 1\right\rangle $ and $\left\vert
2\right\rangle $ as the computational states.

If the TLS is initially in an eigenstate (the ground or excited state) the
coherence of the system will remain zero in free decay. To extract the
relaxation and decoherence times of the SQUID qubit from numerically
simulated time evolution in free decay we assume the initial state of the
qubit is a superposition state with $\rho _{11}\left( 0\right) =\rho
_{12}\left( 0\right) =\rho _{21}\left( 0\right) =\rho _{22}\left( 0\right)
=0.5$. From Eqs. (\ref{i-1}) to (\ref{i-3}), in this case $u_{0}=1$ and $%
v_{0}=w_{0}=0$. Using this initial state one can calculate population and
coherence of the SQUID qubit in free decay by numerically solving the master
equation given by Eqs. (\ref{1}) and (\ref{2}).

In FIG. 4 (a) and (b) the solid lines are the numerical results of the
evolution of population inversion $\left( \rho _{22}-\rho _{11}\right) $ and
squared modulus of coherence $\left\vert \rho _{12}\right\vert ^{2}$,
respectively. Since the coherence is usually a complex and fast oscillating
quantity, we use $\left\vert \rho _{12}\right\vert ^{2}$ instead of $\rho
_{12}$ to estimate the decoherence time. It is shown that both the
population inversion and the squared modulus of coherence undergo simple
exponential decays.

To evaluate the relaxation and decoherence times, we fit the numerically
obtained $\left( \rho _{22}-\rho _{11}\right) $ and $\left\vert \rho
_{12}\right\vert ^{2}$ with exponential functions%
\begin{equation}
\rho _{22}-\rho _{11}=y_{1}+z_{1}e^{-t/\tau _{1}},  \label{f-1}
\end{equation}%
and%
\begin{equation}
\left\vert \rho _{12}\right\vert ^{2}=y_{2}+z_{2}e^{-2t/\tau _{2}},
\label{f-2}
\end{equation}%
respectively. The results of least-square fitting are plotted in FIG. 4 (a)
and (b) with dashed lines, where $\tau _{1}=102.43$ ns and $\tau _{2}=188.27$
ns (and other fitting parameters are $y_{1}=-0.61198$, $z_{1}=0.61198$, $%
y_{2}=0$, and $z_{2}=0.25$). For the weak damping considered here $\theta
_{0}\approx 0$. According to Eqs. (\ref{m30-3}) to (\ref{m-30}), for the
qubit in free decay, the relaxation time is $T_{1}=\tau _{1}$ and the
decoherence time is $T_{2}=\tau _{2}$. Using Eq. (\ref{x-81-1}), the
dephasing time $T_{\varphi }=2324.83$ ns for this special case. Note that
since in this particular case $T_{\varphi }\gg 2T_{1}$ the decoherence time
is limited by the relaxation time.

For comparison, we calculate the relaxation, decoherence, and dephasing
times using the analytical expressions given by Eqs. (\ref{x-81-1-0}) to (%
\ref{x-81-1-2}), Eq. (\ref{i-0-0}), and Eq. (\ref{3-6}). The results are $%
T_{1}=102.43$ ns, $T_{2}=188.27$ ns, and $T_{\varphi }=2324.83$ ns. These
results are exactly the same as the numerical results, demonstrating the
analytical and numerical methods give identical relaxation and decoherence
times for the SQUID qubit in free decay within the relative error of the
least-square fitting.

\subsubsection{Resonantly driven SQUID qubit - Rabi oscillation}

To illustrate the effect of driving field on relaxation and decoherence we
consider a SQUID qubit being in the ground state initially driven by a
resonant microwave field. For the ground state, $\rho _{11}\left( 0\right)
=1 $ and $\rho _{22}\left( 0\right) =\rho _{21}\left( 0\right) =\rho
_{12}\left( 0\right) =0$ and thus $u_{0}=v_{0}=0$ and $w_{0}=1$ from Eqs. (%
\ref{i-1}) to (\ref{i-3}). In this case, from Eq. (\ref{i-0-1}) the
intrinsic decoherence terms of Eqs. (\ref{x-75}) and (\ref{x-76}) vanish.
With this initial state one can calculate the population and coherence of
the microwave-driven SQUID flux qubit by numerically solving the master
equation.

In FIG. 5 (a) and (b) we plot with the solid lines the evolution of
population difference $\left( \rho _{11}-\rho _{22}\right) $ and squared
modulus of coherence $\left\vert \rho _{12}\right\vert ^{2}$, respectively,
for the SQUID flux qubit driven by a microwave with $\phi _{\mu }=1.0\times
10^{-4}$ and $\omega _{\mu }=\omega _{21}=7.22\times 10^{-2}\omega _{LC}$,
where $\omega _{LC}=2.582\times 10^{11}$ rad/s. The damping strength is $%
\eta =8.62\times 10^{-6}\omega _{LC}$ for the SQUID qubit considered here
while the field strength corresponding to the critically-damped regime is $%
\phi _{\mu c}=4.94\times 10^{-7}$ from Eq. (\ref{b4-2}). Thus the SQUID
qubit is in the underdamped regime. As a consequence, both the $\left( \rho
_{11}-\rho _{22}\right) $ and $\left\vert \rho _{12}\right\vert ^{2}$
undergo damped Rabi oscillations with Rabi frequency $\Omega =1.74\times
10^{-3}$ $\omega _{LC}$, as shown in FIG. 5 (a) and (b).

To extract the relaxation and decoherence times of the driven qubit, we fit
the numerically calculated $\left( \rho _{11}-\rho _{22}\right) $ and $%
\left\vert \rho _{12}\right\vert ^{2}$ to exponentially damped oscillating
functions%
\begin{equation}
\rho _{11}-\rho _{22}=\widetilde{y}_{1}+\widetilde{z}_{1}\sin \left( \Omega
t+\varphi _{1}\right) e^{-t/\widetilde{\tau }_{1}},  \label{fit-3}
\end{equation}

and%
\begin{eqnarray}
\left\vert \rho _{12}\right\vert ^{2} &=&\widetilde{y}_{2}+\widetilde{z}%
_{2}\sin \left( \Omega t+\varphi _{2}\right) e^{-t/\widetilde{\tau }_{22}} 
\notag \\
&&+\widetilde{z}_{3}\sin ^{2}\left( \Omega t+\varphi _{2}\right) e^{-2t/%
\widetilde{\tau }_{22}},  \label{fit-4}
\end{eqnarray}%
respectively. The results of best fit are shown in FIG. 5 (a) and (b) with
dashed lines, where $\widetilde{\tau }_{1}=132.68$ ns and $\widetilde{\tau }%
_{22}=132.65$ ns (and other fitting parameters are $\widetilde{y}%
_{1}=0.00033 $, $\widetilde{z}_{1}=0.99976$, $\varphi _{1}=1.56276$, $%
\widetilde{y}_{2}=0.00042$, $\widetilde{z}_{2}=0.00665$, $\widetilde{z}%
_{3}=0.24941$, and $\varphi _{2}=-0.01306$). Since $\Omega \ll \omega _{\mu }
$ the system is in the weak field regime. According to Eqs. (\ref{x-75}) to (%
\ref{x-77}), for the resonantly driven qubit, the relaxation time is $%
\widetilde{T}_{1}=\widetilde{\tau }_{1}$ and the field-dependent decoherence
time is $\widetilde{T}_{2,2}=\widetilde{\tau }_{22}$.

For comparison, we have calculated the relaxation and field-dependent
decoherence times using the analytical expressions given by Eq. (\ref{x-86})
and Eqs. (\ref{x-81-1-0}) to (\ref{x-81-1-2}). The results, $\widetilde{T}%
_{1}=\widetilde{T}_{2,2}=132.68$ ns, are in excellent agreement with the
numerical results. We have also calculated the relaxation and decoherence
times of the underdamped qubit driven by microwaves of different field
strength. The results are given in TABLE \ref{Table 2}. It is shown that
compared to free decay the effect of resonant microwave field is to make the
relaxation time longer but the decoherence time shorter. It is also shown
that the relaxation and field-dependent decoherence times obtained from the
numerical calculation are essentially identical and independent of the field
strength, which accord with the analytical results. Note that as the
microwave field becomes stronger the numerically obtained decoherence and
relaxation times begin to deviate from the analytical results due to strong
field effects.

\begin{table}[htbp] \centering%
\caption{Numerical results of relaxation and decoherence times (ns) of the SQUID qubit in free decay and driven decay.\label{Table
2}}%
\begin{tabular}{ccccc}
\hline\hline
Field strength & \ \ \ \ \ \  & Relaxation time & \ \ \ \ \ \  & Decoherence
time \\ \hline
$0$ &  & $102.43$ &  & $188.27$ \\ 
$1\times 10^{-6}$ &  & $132.68$ &  & $132.68$ \\ 
$5\times 10^{-6}$ &  & $132.68$ &  & $132.68$ \\ 
$1\times 10^{-5}$ &  & $132.68$ &  & $132.67$ \\ 
$5\times 10^{-5}$ &  & $132.68$ &  & $132.67$ \\ 
$1\times 10^{-4}$ &  & $132.68$ &  & $132.65$ \\ 
$2\times 10^{-4}$ &  & $132.70$ &  & $132.49$ \\ \hline\hline
\end{tabular}%
\end{table}%

\section{Conclusion}

In summary, by analytically solving the Bloch equation of a resonantly
driven dissipative TLS under the RWA, the dynamical behavior of a driven
qubit is systemically investigated. It is shown that the driving field has
significant effect on the relaxation and decoherence of a qubit. For a
resonantly driven qubit, the population and coherence undergo more
complicated damped oscillations that in general have more than one
exponential decay terms. Multiple relaxation and decoherence times are thus
required to completely characterize the time evolution of the driven qubit.
It is found that the decoherence of a driven qubit can always be decomposed
into the intrinsic and field-dependent ones. The intrinsic and
field-dependent decoherence times equal to the decoherence time of the qubit
in free decay and relaxation time of the driven qubit, respectively. The
relaxation time $\widetilde{T}_{1}$ and field-dependent decoherence time $%
\widetilde{T}_{2,2}$ of the driven qubit are always in between $T_{1}$ and $%
T_{2}$ and smaller than the lesser of $4T_{1}/3$ and $2T_{\varphi }$. The
analytical expressions for calculation of the relaxation and decoherence
times of the driven qubit are derived. These analytical expressions have
been used to study relaxation and decoherence of the microwave-driven SQUID
qubit. The results are in excellent agreement with those obtained by
numerically solving the master equation, confirming the validity of the
analytical expressions. The relations between the relaxation and decoherence
times of a qubit in free decay and driven damped oscillation are obtained.
These relations can be used to extract the decoherence and dephasing times
of a qubit by measuring its population evolution with free decay and
resonantly driven Rabi oscillation experiments.

\begin{acknowledgments}
This work is supported in part by the NSF (DMR-0325551) and by AFOSR, NSA,
and ARDA through DURINT grant (F49620-01-1-0439).
\end{acknowledgments}

\appendix

\section{The solution of Bloch equation of a dissipative TLS in free decay}

In the absence of ac driving fields (free decay), $\varepsilon =\delta =0$,
and the field-dependent terms in the Bloch equitation (\ref{m-2}) to (\ref%
{m-2-2}) vanish. Usually, the coherence components $u$ and $v$ contain fast
oscillating factors. Hence the first term on the right-hand side of Eq. (\ref%
{m-2-2}) oscillates rapidly. In the rotating-reference which rotates with
the coherence components, the last four terms on the right-hand side of Eq. (%
\ref{m-2}) also oscillate fast. After dropping out all the fast oscillating
terms under the RWA, the Bloch equation is approximated to%
\begin{eqnarray}
\frac{du}{dt} &=&\left( R_{12,12}+R_{12,21}\right) u+\omega _{21}v,
\label{m-8} \\
\frac{dv}{dt} &=&-\omega _{21}u+\left( R_{12,12}-R_{12,21}\right) v,
\label{m-8-1} \\
\frac{dw}{dt} &=&-\left( R_{22,11}+R_{11,22}\right) w+\left(
R_{11,22}-R_{22,11}\right) .  \label{m-8-2}
\end{eqnarray}%
Obviously, there is no coupling between the population and coherence in this
special case. Because of this the population and coherence evolve
independently.

If the initial value of the Bloch vector is $\left( u_{0},v_{0},w_{0}\right) 
$, the solution of Eqs. (\ref{m-8}) to (\ref{m-8-2}) is then given by Eqs. (%
\ref{m30-3}) to (\ref{m-30}). The parameters and constants in the solution
are given by%
\begin{equation}
w_{\infty }=\frac{R_{11,22}-R_{22,11}}{R_{22,11}+R_{11,22}},  \label{i-4-4}
\end{equation}%
\begin{equation}
\varpi =\sqrt{\omega _{21}^{2}-R_{12,21}^{2}},  \label{i-4-5}
\end{equation}%
\begin{equation}
\theta _{0}=2\tan ^{-1}\sqrt{\frac{\omega _{21}+R_{12,21}}{\omega
_{21}-R_{12,21}}}-\frac{\pi }{2},  \label{i-4-6}
\end{equation}%
\begin{equation}
A=\frac{u_{0}}{\cos \theta _{0}},  \label{i-0-1}
\end{equation}%
\begin{equation}
B=\frac{v_{0}}{\cos \theta _{0}},  \label{i-0-2}
\end{equation}%
and%
\begin{equation}
D=w_{0}-w_{\infty }.  \label{i-4-3}
\end{equation}

\section{Solutions of Bloch equation of a resonantly driven dissipative TLS}

For a driven dissipative TLS, $\varepsilon \neq 0$ and also $\delta \neq 0$
in general, as will be shown in section V. In this case, we assume that the
trial solution of the Bloch equation (\ref{m-2}) to (\ref{m-2-2}) still has
the form of Eqs. (\ref{m30-3}) to (\ref{m-30}) but with the constants $A$, $%
B $, and $D$ replaced by time-dependent variables $\mu $, $\nu $, and $%
\lambda $. Substituting the trial solution into Eqs. (\ref{m-2}) to (\ref%
{m-2-2}) we obtain three equations with respect to the variables $\mu $, $%
\nu $, and $\lambda $. In the case of the resonant driving field with $%
\omega _{\mu }=\varpi $, using the RWA, these equations are simplified to
the form%
\begin{eqnarray}
\frac{d\mu }{dt} &=&0,  \label{o-11} \\
\frac{d\nu }{dt} &=&-\frac{\varepsilon }{\cos \left( \theta _{0}\right) }%
\left( \lambda e^{-2\eta t}+w_{\infty }e^{\kappa t}\right) ,  \label{o-12} \\
\frac{d\lambda }{dt} &=&\nu \varepsilon \cos \left( \theta _{0}\right)
e^{2\eta t},  \label{o-13}
\end{eqnarray}%
where, $\eta =\left( \gamma -\kappa \right) /2$ is the damping strength.
Eqs. (\ref{o-12}) and (\ref{o-13}) indicate that due to the driving field
the population and coherence are coupled to each other. \cite{Grifoni1999}
Due to this coupling they no longer evolve independently.

The form of the solution of Eqs. (\ref{o-11}) to (\ref{o-13}) depends on the
relative value of $\varepsilon $ to $\eta $. According to this relative
value the dynamics of the driven dissipative TLS can be categorized into
three regimes: underdamped regime when $\varepsilon >\eta $,
critically-damped regime when $\varepsilon =\eta $, and overdamped regime
when $\varepsilon <\eta $. For the underdamped regime with $\varepsilon
>\eta $, the most important scheme for quantum computation, the solution of
Bloch equation is given by Eqs. (\ref{x-75}) to (\ref{x-77}). The parameters
and constants in the solution are given by

\begin{equation}
B_{0}=-\allowbreak \frac{\chi \varepsilon }{\cos \theta _{0}},  \label{a-5}
\end{equation}%
\begin{equation}
B_{1}=\frac{\left\vert w_{0}-\kappa \chi \right\vert }{\cos \left( \theta
_{0}\right) }\sqrt{1+\frac{1}{\Omega ^{2}}\left( \eta -\varepsilon \frac{%
v_{0}+\allowbreak \varepsilon \chi }{w_{0}-\kappa \chi }\right) ^{2}},
\label{a-5-0-1}
\end{equation}%
\begin{equation}
D_{0}=\kappa \chi ,  \label{a-5-0}
\end{equation}%
\begin{equation}
D_{1}=B_{1}\cos \theta _{0},  \label{a-5-0-2}
\end{equation}%
\begin{equation}
\theta _{1}=\theta _{2}+\theta _{3},  \label{a-5-1}
\end{equation}%
\begin{equation}
\theta _{2}=\tan ^{-1}\frac{\Omega \left( w_{0}-\kappa \chi \right) }{%
\varepsilon v_{0}-\eta w_{0}+\left( \eta \kappa +\allowbreak \varepsilon
^{2}\right) \chi },  \label{a-7}
\end{equation}%
\begin{equation}
\theta _{3}=\tan ^{-1}\frac{\Omega }{\eta },  \label{a-6}
\end{equation}%
and 
\begin{equation}
\chi =\frac{\gamma w_{\infty }}{\varepsilon ^{2}+\kappa \gamma }.
\label{x-7}
\end{equation}
For the overdamped regime with $\varepsilon <\eta $, the solution of Bloch
equation can be obtained by replacing $\Omega \ $with $i\sqrt{\eta
^{2}-\varepsilon ^{2}}$ in Eqs. (\ref{x-75}) to (\ref{x-77}). In this case,
the population difference and coherence components evolve with more than one
exponential decaying terms and the coherence components are composed of
intrinsic and field-dependent decoherences. For the critically-damped regime
with $\varepsilon =\eta $, the solution of Bloch equation can be obtained by
setting $\Omega =0$ in Eqs. (\ref{x-75}) to (\ref{x-77}). In this case, the
system undergoes a nonexponential decay owing to the nonexponential decay
factor $te^{-\Gamma t}$ in the population difference and coherence
components. Note that the analytical solutions obtained here have been
confirmed by the \emph{matrix exponent method}. \cite{Alekseev1992}

\section{Equivalent admittance of the external circuit}

In this paper we assume that the circuit denoted by $Y_{e}(\omega )$ in FIG.
1 (a) is an $RC$ circuit with effective admittance $Y_{e}(\omega )$ given by%
\begin{equation}
Y_{e}(\omega )=\frac{1}{R_{0}}+\frac{1}{R_{e}+1/j\omega C_{e}}.
\label{3-0-0}
\end{equation}%
The corresponding impedance $Z_{e}$ is given by%
\begin{equation}
Z_{e}(\omega )=\frac{1}{Y_{e}(\omega )}=R_{eff}(\omega )+\frac{1}{j\omega
C_{eff}(\omega )},  \label{3-0}
\end{equation}%
where, $R_{eff}$ and $C_{eff}$ are the effective resistance and capacity
given by%
\begin{equation}
R_{eff}(\omega )=\frac{R_{0}+\omega ^{2}C_{e}^{2}R_{e}R_{0}\left(
R_{e}+R_{0}\allowbreak \right) }{1+\omega ^{2}C_{e}^{2}\left(
R_{e}+R_{0}\right) ^{2}},  \label{4}
\end{equation}%
and%
\begin{equation}
C_{eff}(\omega )=C_{e}\left[ \frac{1}{\omega ^{2}R_{0}^{2}C_{e}^{2}}+\left( 
\frac{R_{e}+R_{0}}{R_{0}}\right) ^{2}\right] ,  \label{5}
\end{equation}%
respectively. They are functions of $\omega $. When $R_{0}\allowbreak
\rightarrow \infty $, $R_{eff}=R_{e}$ and $C_{eff}=C_{e}$.

For the external circuit shown in FIG. 1 (a), the circuit equations are%
\begin{eqnarray}
V_{1} &=&j\omega LI_{1}-j\omega MI_{2},  \label{A-a7} \\
0 &=&-j\omega MI_{1}+\left( j\omega L_{e}+R_{eff}+\frac{1}{j\omega C_{eff}}%
\right) I_{2}.  \label{A-a7-1}
\end{eqnarray}%
The equivalent impedance $Z\left( \omega \right) $ is therefore calculated
from%
\begin{equation}
Z\left( \omega \right) =\frac{V_{1}}{I_{1}}=j\omega L+\frac{\omega ^{2}M^{2}%
}{j\omega L_{e}+R_{eff}+1/j\omega C_{eff}}.  \label{A-a8}
\end{equation}%
The equivalent admittance $Y\left( \omega \right) $ can be calculated from
the equivalent impedance $Z\left( \omega \right) $ by $Y\left( \omega
\right) =1/Z\left( \omega \right) $. The real part of the equivalent
admittance, $Y_{R}\left( \omega \right) $, is given by Eq. (\ref{3-6}),
where,%
\begin{equation}
F_{0}\left( \omega \right) =\frac{M^{2}R_{eff}}{\varsigma ^{2}},
\label{A-a10}
\end{equation}%
\begin{equation}
G_{0}\left( \omega \right) =\frac{2L}{\varsigma C_{eff}}+\frac{L^{2}}{%
\varsigma ^{2}}\left( R_{eff}^{2}+\frac{1}{\omega ^{2}C_{eff}^{2}}\right) ,
\label{A-a11}
\end{equation}%
and%
\begin{equation}
\varsigma =M^{2}-LL_{e}.  \label{A-a14}
\end{equation}%
Eq. (\ref{3-6}) shows that $Y_{R}\left( \omega \right) $ is an even function
of $\omega $. It is also the case for most of superconducting qubits.

\bibliographystyle{apsrev}
\bibliography{noise1}

\begin{thebibliography}{69}
\expandafter\ifx\csname natexlab\endcsname\relax\def\natexlab#1{#1}\fi
\expandafter\ifx\csname bibnamefont\endcsname\relax
  \def\bibnamefont#1{#1}\fi
\expandafter\ifx\csname bibfnamefont\endcsname\relax
  \def\bibfnamefont#1{#1}\fi
\expandafter\ifx\csname citenamefont\endcsname\relax
  \def\citenamefont#1{#1}\fi
\expandafter\ifx\csname url\endcsname\relax
  \def\url#1{\texttt{#1}}\fi
\expandafter\ifx\csname urlprefix\endcsname\relax\def\urlprefix{URL }\fi
\providecommand{\bibinfo}[2]{#2}
\providecommand{\eprint}[2][]{\url{#2}}

\bibitem[{\citenamefont{Nakamura et~al.}(1999)\citenamefont{Nakamura, {Yu. A.
  Pashkin}, and Tsai}}]{Nakamura1999}
\bibinfo{author}{\bibfnamefont{Y.}~\bibnamefont{Nakamura}},
  \bibinfo{author}{\bibnamefont{{Yu. A. Pashkin}}}, \bibnamefont{and}
  \bibinfo{author}{\bibfnamefont{J.~S.} \bibnamefont{Tsai}},
  \bibinfo{journal}{Nature (London)} \textbf{\bibinfo{volume}{398}},
  \bibinfo{pages}{786} (\bibinfo{year}{1999}).

\bibitem[{\citenamefont{Vion et~al.}(2002)\citenamefont{Vion, Aassime, Cottet,
  Joyez, Pothier, Urbina, Esteve, and Devoret}}]{Vion2002}
\bibinfo{author}{\bibfnamefont{D.}~\bibnamefont{Vion}},
  \bibinfo{author}{\bibfnamefont{A.}~\bibnamefont{Aassime}},
  \bibinfo{author}{\bibfnamefont{A.}~\bibnamefont{Cottet}},
  \bibinfo{author}{\bibfnamefont{P.}~\bibnamefont{Joyez}},
  \bibinfo{author}{\bibfnamefont{H.}~\bibnamefont{Pothier}},
  \bibinfo{author}{\bibfnamefont{C.}~\bibnamefont{Urbina}},
  \bibinfo{author}{\bibfnamefont{D.}~\bibnamefont{Esteve}}, \bibnamefont{and}
  \bibinfo{author}{\bibfnamefont{M.~H.} \bibnamefont{Devoret}},
  \bibinfo{journal}{Science} \textbf{\bibinfo{volume}{296}},
  \bibinfo{pages}{886} (\bibinfo{year}{2002}).

\bibitem[{\citenamefont{Yu et~al.}(2002)\citenamefont{Yu, Han, Chu, Chu, and
  Wang}}]{Yu2002}
\bibinfo{author}{\bibfnamefont{Y.}~\bibnamefont{Yu}},
  \bibinfo{author}{\bibfnamefont{S.}~\bibnamefont{Han}},
  \bibinfo{author}{\bibfnamefont{X.}~\bibnamefont{Chu}},
  \bibinfo{author}{\bibfnamefont{S.-I.} \bibnamefont{Chu}}, \bibnamefont{and}
  \bibinfo{author}{\bibfnamefont{Z.}~\bibnamefont{Wang}},
  \bibinfo{journal}{Science} \textbf{\bibinfo{volume}{296}},
  \bibinfo{pages}{889} (\bibinfo{year}{2002}).

\bibitem[{\citenamefont{Martinis et~al.}(2002)\citenamefont{Martinis, Nam,
  Aumentado, and Urbina}}]{Martinis2002}
\bibinfo{author}{\bibfnamefont{J.~M.} \bibnamefont{Martinis}},
  \bibinfo{author}{\bibfnamefont{S.}~\bibnamefont{Nam}},
  \bibinfo{author}{\bibfnamefont{J.}~\bibnamefont{Aumentado}},
  \bibnamefont{and} \bibinfo{author}{\bibfnamefont{C.}~\bibnamefont{Urbina}},
  \bibinfo{journal}{Phys. Rev. Lett.} \textbf{\bibinfo{volume}{89}},
  \bibinfo{pages}{117901} (\bibinfo{year}{2002}).

\bibitem[{\citenamefont{Friedman et~al.}(2000)\citenamefont{Friedman, Patel,
  Chen, Tolpygo, and Lukens}}]{Friedman2000}
\bibinfo{author}{\bibfnamefont{J.~R.} \bibnamefont{Friedman}},
  \bibinfo{author}{\bibfnamefont{V.}~\bibnamefont{Patel}},
  \bibinfo{author}{\bibfnamefont{W.}~\bibnamefont{Chen}},
  \bibinfo{author}{\bibfnamefont{S.~K.} \bibnamefont{Tolpygo}},
  \bibnamefont{and} \bibinfo{author}{\bibfnamefont{J.~E.}
  \bibnamefont{Lukens}}, \bibinfo{journal}{Nature (London)}
  \textbf{\bibinfo{volume}{406}}, \bibinfo{pages}{43} (\bibinfo{year}{2000}).

\bibitem[{\citenamefont{{van der Wal} et~al.}(2000)\citenamefont{{van der Wal},
  {ter Haar}, Wilhelm, Schouten, Harmans, Orlando, Lloyd, and Mooij}}]{Wal2000}
\bibinfo{author}{\bibfnamefont{C.~H.} \bibnamefont{{van der Wal}}},
  \bibinfo{author}{\bibfnamefont{A.~C.~J.} \bibnamefont{{ter Haar}}},
  \bibinfo{author}{\bibfnamefont{F.~K.} \bibnamefont{Wilhelm}},
  \bibinfo{author}{\bibfnamefont{R.~N.} \bibnamefont{Schouten}},
  \bibinfo{author}{\bibfnamefont{C.~J. P.~M.} \bibnamefont{Harmans}},
  \bibinfo{author}{\bibfnamefont{T.~P.} \bibnamefont{Orlando}},
  \bibinfo{author}{\bibfnamefont{S.}~\bibnamefont{Lloyd}}, \bibnamefont{and}
  \bibinfo{author}{\bibfnamefont{J.~E.} \bibnamefont{Mooij}},
  \bibinfo{journal}{Science} \textbf{\bibinfo{volume}{290}},
  \bibinfo{pages}{773} (\bibinfo{year}{2000}).

\bibitem[{\citenamefont{Chiorescu et~al.}(2003)\citenamefont{Chiorescu,
  Nakamura, Harmans, and Mooij}}]{Chiorescu2003}
\bibinfo{author}{\bibfnamefont{I.}~\bibnamefont{Chiorescu}},
  \bibinfo{author}{\bibfnamefont{Y.}~\bibnamefont{Nakamura}},
  \bibinfo{author}{\bibfnamefont{C.~J. P.~M.} \bibnamefont{Harmans}},
  \bibnamefont{and} \bibinfo{author}{\bibfnamefont{J.~E.} \bibnamefont{Mooij}},
  \bibinfo{journal}{Science} \textbf{\bibinfo{volume}{299}},
  \bibinfo{pages}{1869} (\bibinfo{year}{2003}).

\bibitem[{\citenamefont{Chiorescu et~al.}(2004)\citenamefont{Chiorescu, Bertet,
  Semba, Nakamura, Harmans, and Mooij}}]{Chiorescu2004}
\bibinfo{author}{\bibfnamefont{I.}~\bibnamefont{Chiorescu}},
  \bibinfo{author}{\bibfnamefont{P.}~\bibnamefont{Bertet}},
  \bibinfo{author}{\bibfnamefont{K.}~\bibnamefont{Semba}},
  \bibinfo{author}{\bibfnamefont{Y.}~\bibnamefont{Nakamura}},
  \bibinfo{author}{\bibfnamefont{C.~J. P.~M.} \bibnamefont{Harmans}},
  \bibnamefont{and} \bibinfo{author}{\bibfnamefont{J.~E.} \bibnamefont{Mooij}},
  \bibinfo{journal}{Nature} \textbf{\bibinfo{volume}{431}},
  \bibinfo{pages}{159} (\bibinfo{year}{2004}).

\bibitem[{\citenamefont{{Yu. A. Pashkin} et~al.}(2003)\citenamefont{{Yu. A.
  Pashkin}, Yamamoto, Astafiev, Nakamura, Averin, and Tsai}}]{Pashkin2003}
\bibinfo{author}{\bibnamefont{{Yu. A. Pashkin}}},
  \bibinfo{author}{\bibfnamefont{T.}~\bibnamefont{Yamamoto}},
  \bibinfo{author}{\bibfnamefont{O.}~\bibnamefont{Astafiev}},
  \bibinfo{author}{\bibfnamefont{Y.}~\bibnamefont{Nakamura}},
  \bibinfo{author}{\bibfnamefont{D.~V.} \bibnamefont{Averin}},
  \bibnamefont{and} \bibinfo{author}{\bibfnamefont{J.~S.} \bibnamefont{Tsai}},
  \bibinfo{journal}{Nature (London)} \textbf{\bibinfo{volume}{421}},
  \bibinfo{pages}{823} (\bibinfo{year}{2003}).

\bibitem[{\citenamefont{Yamamoto et~al.}(2003)\citenamefont{Yamamoto, {Yu. A.
  Pashkin}, Astafiev, Nakamura, and Tsai}}]{Yamamoto2003}
\bibinfo{author}{\bibfnamefont{T.}~\bibnamefont{Yamamoto}},
  \bibinfo{author}{\bibnamefont{{Yu. A. Pashkin}}},
  \bibinfo{author}{\bibfnamefont{O.}~\bibnamefont{Astafiev}},
  \bibinfo{author}{\bibfnamefont{Y.}~\bibnamefont{Nakamura}}, \bibnamefont{and}
  \bibinfo{author}{\bibfnamefont{J.~S.} \bibnamefont{Tsai}},
  \bibinfo{journal}{Nature (London)} \textbf{\bibinfo{volume}{425}},
  \bibinfo{pages}{941} (\bibinfo{year}{2003}).

\bibitem[{\citenamefont{Berkley
  et~al.}(2003{\natexlab{a}})\citenamefont{Berkley, Xu, Ramos, Gubrud, Strauch,
  Johnson, Anderson, Dragt, Lobb, and Wellstood}}]{Berkley2003}
\bibinfo{author}{\bibfnamefont{A.~J.} \bibnamefont{Berkley}},
  \bibinfo{author}{\bibfnamefont{H.}~\bibnamefont{Xu}},
  \bibinfo{author}{\bibfnamefont{R.~C.} \bibnamefont{Ramos}},
  \bibinfo{author}{\bibfnamefont{M.~A.} \bibnamefont{Gubrud}},
  \bibinfo{author}{\bibfnamefont{F.~W.} \bibnamefont{Strauch}},
  \bibinfo{author}{\bibfnamefont{P.~R.} \bibnamefont{Johnson}},
  \bibinfo{author}{\bibfnamefont{J.~R.} \bibnamefont{Anderson}},
  \bibinfo{author}{\bibfnamefont{A.~J.} \bibnamefont{Dragt}},
  \bibinfo{author}{\bibfnamefont{C.~J.} \bibnamefont{Lobb}}, \bibnamefont{and}
  \bibinfo{author}{\bibfnamefont{F.~C.} \bibnamefont{Wellstood}},
  \bibinfo{journal}{Science} \textbf{\bibinfo{volume}{300}},
  \bibinfo{pages}{1548} (\bibinfo{year}{2003}{\natexlab{a}}).

\bibitem[{\citenamefont{McDermott et~al.}(2005)\citenamefont{McDermott,
  Simmonds, Steffen, Cooper, Cicak, Osborn, Oh, Pappas, and
  Martinis}}]{McDermott05}
\bibinfo{author}{\bibfnamefont{R.}~\bibnamefont{McDermott}},
  \bibinfo{author}{\bibfnamefont{R.~W.} \bibnamefont{Simmonds}},
  \bibinfo{author}{\bibfnamefont{M.}~\bibnamefont{Steffen}},
  \bibinfo{author}{\bibfnamefont{K.~B.} \bibnamefont{Cooper}},
  \bibinfo{author}{\bibfnamefont{K.}~\bibnamefont{Cicak}},
  \bibinfo{author}{\bibfnamefont{K.~D.} \bibnamefont{Osborn}},
  \bibinfo{author}{\bibfnamefont{S.}~\bibnamefont{Oh}},
  \bibinfo{author}{\bibfnamefont{D.~P.} \bibnamefont{Pappas}},
  \bibnamefont{and} \bibinfo{author}{\bibfnamefont{J.~M.}
  \bibnamefont{Martinis}}, \bibinfo{journal}{Science}
  \textbf{\bibinfo{volume}{307}}, \bibinfo{pages}{1299} (\bibinfo{year}{2005}).

\bibitem[{\citenamefont{Astafiev et~al.}(2004)\citenamefont{Astafiev, {Yu. A.
  Pashkin}, Nakamura, Yamamoto, and Tsai}}]{Astafiev04}
\bibinfo{author}{\bibfnamefont{O.}~\bibnamefont{Astafiev}},
  \bibinfo{author}{\bibnamefont{{Yu. A. Pashkin}}},
  \bibinfo{author}{\bibfnamefont{Y.}~\bibnamefont{Nakamura}},
  \bibinfo{author}{\bibfnamefont{T.}~\bibnamefont{Yamamoto}}, \bibnamefont{and}
  \bibinfo{author}{\bibfnamefont{J.~S.} \bibnamefont{Tsai}},
  \bibinfo{journal}{Phys. Rev. Lett.} \textbf{\bibinfo{volume}{93}},
  \bibinfo{pages}{267007} (\bibinfo{year}{2004}).

\bibitem[{\citenamefont{{G. Ithier, E. Collin, P. Joyez, P. J. Meeson, D. Vion,
  D. Esteve, F. Chiarello, A. Shnirman, Y. Makhlin, J. Schriefl, and G.
  Sch\"{o}n}}(2005)}]{Ithier05}
\bibinfo{author}{\bibnamefont{{G. Ithier, E. Collin, P. Joyez, P. J. Meeson, D.
  Vion, D. Esteve, F. Chiarello, A. Shnirman, Y. Makhlin, J. Schriefl, and G.
  Sch\"{o}n}}}, \bibinfo{journal}{Phys. Rev. B} \textbf{\bibinfo{volume}{72}},
  \bibinfo{pages}{134519} (\bibinfo{year}{2005}).

\bibitem[{\citenamefont{Makhlin et~al.}(2001)\citenamefont{Makhlin, Schon, and
  Shnirman}}]{Makhlin2001}
\bibinfo{author}{\bibfnamefont{Y.}~\bibnamefont{Makhlin}},
  \bibinfo{author}{\bibfnamefont{G.}~\bibnamefont{Schon}}, \bibnamefont{and}
  \bibinfo{author}{\bibfnamefont{A.}~\bibnamefont{Shnirman}},
  \bibinfo{journal}{Rev. Mod. Phys.} \textbf{\bibinfo{volume}{73}},
  \bibinfo{pages}{357} (\bibinfo{year}{2001}).

\bibitem[{\citenamefont{Mooij et~al.}(1999)\citenamefont{Mooij, Orlando,
  Levitov, Tian, {van der Wal}, and Lloyd}}]{Mooij1999}
\bibinfo{author}{\bibfnamefont{J.~E.} \bibnamefont{Mooij}},
  \bibinfo{author}{\bibfnamefont{T.~P.} \bibnamefont{Orlando}},
  \bibinfo{author}{\bibfnamefont{L.}~\bibnamefont{Levitov}},
  \bibinfo{author}{\bibfnamefont{L.}~\bibnamefont{Tian}},
  \bibinfo{author}{\bibfnamefont{C.~H.} \bibnamefont{{van der Wal}}},
  \bibnamefont{and} \bibinfo{author}{\bibfnamefont{S.}~\bibnamefont{Lloyd}},
  \bibinfo{journal}{Science} \textbf{\bibinfo{volume}{285}},
  \bibinfo{pages}{1036} (\bibinfo{year}{1999}).

\bibitem[{\citenamefont{Burkard et~al.}(2005)\citenamefont{Burkard, DiVincenzo,
  Bertet, Chiorescu, and Mooij}}]{Burkard05-1}
\bibinfo{author}{\bibfnamefont{G.}~\bibnamefont{Burkard}},
  \bibinfo{author}{\bibfnamefont{D.~P.} \bibnamefont{DiVincenzo}},
  \bibinfo{author}{\bibfnamefont{P.}~\bibnamefont{Bertet}},
  \bibinfo{author}{\bibfnamefont{I.}~\bibnamefont{Chiorescu}},
  \bibnamefont{and} \bibinfo{author}{\bibfnamefont{J.~E.} \bibnamefont{Mooij}},
  \bibinfo{journal}{Phys. Rev. B} \textbf{\bibinfo{volume}{71}},
  \bibinfo{pages}{134504} (\bibinfo{year}{2005}).

\bibitem[{\citenamefont{Goorden et~al.}(2004)\citenamefont{Goorden, Thorwart,
  and Grifoni}}]{Goorden04}
\bibinfo{author}{\bibfnamefont{M.~C.} \bibnamefont{Goorden}},
  \bibinfo{author}{\bibfnamefont{M.}~\bibnamefont{Thorwart}}, \bibnamefont{and}
  \bibinfo{author}{\bibfnamefont{M.}~\bibnamefont{Grifoni}},
  \bibinfo{journal}{Phys. Rev. Lett.} \textbf{\bibinfo{volume}{93}},
  \bibinfo{pages}{267005} (\bibinfo{year}{2004}).

\bibitem[{\citenamefont{Makhlin and Shnirman}(2004)}]{Makhlin04}
\bibinfo{author}{\bibfnamefont{Y.}~\bibnamefont{Makhlin}} \bibnamefont{and}
  \bibinfo{author}{\bibfnamefont{A.}~\bibnamefont{Shnirman}},
  \bibinfo{journal}{Phys. Rev. Lett.} \textbf{\bibinfo{volume}{92}},
  \bibinfo{pages}{178301} (\bibinfo{year}{2004}).

\bibitem[{\citenamefont{Orlando et~al.}(2002)\citenamefont{Orlando, Tian,
  Crankshaw, Lloyd, {van der Wal}, Mooij, and Wilhelm}}]{Orlando2002}
\bibinfo{author}{\bibfnamefont{T.~P.} \bibnamefont{Orlando}},
  \bibinfo{author}{\bibfnamefont{L.}~\bibnamefont{Tian}},
  \bibinfo{author}{\bibfnamefont{D.~S.} \bibnamefont{Crankshaw}},
  \bibinfo{author}{\bibfnamefont{S.}~\bibnamefont{Lloyd}},
  \bibinfo{author}{\bibfnamefont{C.~H.} \bibnamefont{{van der Wal}}},
  \bibinfo{author}{\bibfnamefont{J.}~\bibnamefont{Mooij}}, \bibnamefont{and}
  \bibinfo{author}{\bibfnamefont{F.}~\bibnamefont{Wilhelm}},
  \bibinfo{journal}{Physica C} \textbf{\bibinfo{volume}{368}},
  \bibinfo{pages}{294} (\bibinfo{year}{2002}).

\bibitem[{\citenamefont{{Van Harlingen} et~al.}(2004)\citenamefont{{Van
  Harlingen}, Robertson, Plourde, Reichardt, Crane, and Clarke}}]{Harlingen04}
\bibinfo{author}{\bibfnamefont{D.~J.} \bibnamefont{{Van Harlingen}}},
  \bibinfo{author}{\bibfnamefont{T.~L.} \bibnamefont{Robertson}},
  \bibinfo{author}{\bibfnamefont{B.~L.~T.} \bibnamefont{Plourde}},
  \bibinfo{author}{\bibfnamefont{P.~A.} \bibnamefont{Reichardt}},
  \bibinfo{author}{\bibfnamefont{T.~A.} \bibnamefont{Crane}}, \bibnamefont{and}
  \bibinfo{author}{\bibfnamefont{J.}~\bibnamefont{Clarke}},
  \bibinfo{journal}{Phys. Rev. Rev. B} \textbf{\bibinfo{volume}{70}},
  \bibinfo{pages}{064517} (\bibinfo{year}{2004}).

\bibitem[{\citenamefont{Zhou et~al.}(2004)\citenamefont{Zhou, Chu, and
  Han}}]{Zhou2004}
\bibinfo{author}{\bibfnamefont{Z.}~\bibnamefont{Zhou}},
  \bibinfo{author}{\bibfnamefont{S.-I.} \bibnamefont{Chu}}, \bibnamefont{and}
  \bibinfo{author}{\bibfnamefont{S.}~\bibnamefont{Han}},
  \bibinfo{journal}{Phys. Rev. B} \textbf{\bibinfo{volume}{70}},
  \bibinfo{pages}{094513} (\bibinfo{year}{2004}).

\bibitem[{\citenamefont{Shresta et~al.}(2005)\citenamefont{Shresta,
  Anastopoulos, Dragulescu, and Hu}}]{Shresta05}
\bibinfo{author}{\bibfnamefont{S.}~\bibnamefont{Shresta}},
  \bibinfo{author}{\bibfnamefont{C.}~\bibnamefont{Anastopoulos}},
  \bibinfo{author}{\bibfnamefont{A.}~\bibnamefont{Dragulescu}},
  \bibnamefont{and} \bibinfo{author}{\bibfnamefont{B.~L.} \bibnamefont{Hu}},
  \bibinfo{journal}{Phys. Rev. A} \textbf{\bibinfo{volume}{71}},
  \bibinfo{pages}{022109} (\bibinfo{year}{2005}).

\bibitem[{\citenamefont{Anastopoulos and Hu}(2000)}]{Anastopoulos2000}
\bibinfo{author}{\bibfnamefont{C.}~\bibnamefont{Anastopoulos}}
  \bibnamefont{and} \bibinfo{author}{\bibfnamefont{B.~L.} \bibnamefont{Hu}},
  \bibinfo{journal}{Phys. Rev. A} \textbf{\bibinfo{volume}{62}},
  \bibinfo{pages}{033821} (\bibinfo{year}{2000}).

\bibitem[{\citenamefont{Robertson et~al.}(2005)\citenamefont{Robertson,
  Plourde, Hime, Linzen, Reichardt, Wilhelm, and Clarke}}]{Robertson05}
\bibinfo{author}{\bibfnamefont{T.~L.} \bibnamefont{Robertson}},
  \bibinfo{author}{\bibfnamefont{B.~L.~T.} \bibnamefont{Plourde}},
  \bibinfo{author}{\bibfnamefont{T.}~\bibnamefont{Hime}},
  \bibinfo{author}{\bibfnamefont{S.}~\bibnamefont{Linzen}},
  \bibinfo{author}{\bibfnamefont{P.~A.} \bibnamefont{Reichardt}},
  \bibinfo{author}{\bibfnamefont{F.~K.} \bibnamefont{Wilhelm}},
  \bibnamefont{and} \bibinfo{author}{\bibfnamefont{J.}~\bibnamefont{Clarke}},
  \bibinfo{journal}{Phys. Rev. B} \textbf{\bibinfo{volume}{72}},
  \bibinfo{pages}{024513} (\bibinfo{year}{2005}).

\bibitem[{\citenamefont{Cheng and Silbey}(2004)}]{Cheng2004}
\bibinfo{author}{\bibfnamefont{Y.~C.} \bibnamefont{Cheng}} \bibnamefont{and}
  \bibinfo{author}{\bibfnamefont{R.~J.} \bibnamefont{Silbey}},
  \bibinfo{journal}{Phys. Rev. A} \textbf{\bibinfo{volume}{69}},
  \bibinfo{pages}{052325} (\bibinfo{year}{2004}).

\bibitem[{\citenamefont{Storcz and Wilhelm}(2003)}]{Storcz2003}
\bibinfo{author}{\bibfnamefont{M.~J.} \bibnamefont{Storcz}} \bibnamefont{and}
  \bibinfo{author}{\bibfnamefont{F.~K.} \bibnamefont{Wilhelm}},
  \bibinfo{journal}{Phys. Rev. A} \textbf{\bibinfo{volume}{67}},
  \bibinfo{pages}{042319} (\bibinfo{year}{2003}).

\bibitem[{\citenamefont{Thorwart and H\"{a}nggi}(2001)}]{Thorwart2001}
\bibinfo{author}{\bibfnamefont{M.}~\bibnamefont{Thorwart}} \bibnamefont{and}
  \bibinfo{author}{\bibfnamefont{P.}~\bibnamefont{H\"{a}nggi}},
  \bibinfo{journal}{Phys. Rev. A} \textbf{\bibinfo{volume}{65}},
  \bibinfo{pages}{012309} (\bibinfo{year}{2001}).

\bibitem[{\citenamefont{Governale et~al.}(2001)\citenamefont{Governale,
  Grifoni, and Schön}}]{Governale2001}
\bibinfo{author}{\bibfnamefont{M.}~\bibnamefont{Governale}},
  \bibinfo{author}{\bibfnamefont{M.}~\bibnamefont{Grifoni}}, \bibnamefont{and}
  \bibinfo{author}{\bibfnamefont{G.}~\bibnamefont{Schön}},
  \bibinfo{journal}{Chem. Phys.} \textbf{\bibinfo{volume}{268}},
  \bibinfo{pages}{273} (\bibinfo{year}{2001}).

\bibitem[{\citenamefont{Tian et~al.}(2002)\citenamefont{Tian, Lloyd, and
  Orlando}}]{Tian2002}
\bibinfo{author}{\bibfnamefont{L.}~\bibnamefont{Tian}},
  \bibinfo{author}{\bibfnamefont{S.}~\bibnamefont{Lloyd}}, \bibnamefont{and}
  \bibinfo{author}{\bibfnamefont{T.~P.} \bibnamefont{Orlando}},
  \bibinfo{journal}{Phys. Rev. B} \textbf{\bibinfo{volume}{65}},
  \bibinfo{pages}{144516} (\bibinfo{year}{2002}).

\bibitem[{\citenamefont{Xu et~al.}(2005)\citenamefont{Xu, Berkley, Ramos,
  Gubrud, Johnson, Strauch, Dragt, Anderson, Lobb, and Wellstood}}]{Xu2005}
\bibinfo{author}{\bibfnamefont{H.}~\bibnamefont{Xu}},
  \bibinfo{author}{\bibfnamefont{A.~J.} \bibnamefont{Berkley}},
  \bibinfo{author}{\bibfnamefont{R.~C.} \bibnamefont{Ramos}},
  \bibinfo{author}{\bibfnamefont{M.~A.} \bibnamefont{Gubrud}},
  \bibinfo{author}{\bibfnamefont{P.~R.} \bibnamefont{Johnson}},
  \bibinfo{author}{\bibfnamefont{F.~W.} \bibnamefont{Strauch}},
  \bibinfo{author}{\bibfnamefont{A.~J.} \bibnamefont{Dragt}},
  \bibinfo{author}{\bibfnamefont{J.~R.} \bibnamefont{Anderson}},
  \bibinfo{author}{\bibfnamefont{C.~J.} \bibnamefont{Lobb}}, \bibnamefont{and}
  \bibinfo{author}{\bibfnamefont{F.~C.} \bibnamefont{Wellstood}},
  \bibinfo{journal}{Phys. Rev. B} \textbf{\bibinfo{volume}{71}},
  \bibinfo{pages}{064512} (\bibinfo{year}{2005}).

\bibitem[{\citenamefont{Hartmann et~al.}(2000)\citenamefont{Hartmann, Goychuk,
  Grifoni, and H\"{a}nggi}}]{Hartmann2000}
\bibinfo{author}{\bibfnamefont{L.}~\bibnamefont{Hartmann}},
  \bibinfo{author}{\bibfnamefont{I.}~\bibnamefont{Goychuk}},
  \bibinfo{author}{\bibfnamefont{M.}~\bibnamefont{Grifoni}}, \bibnamefont{and}
  \bibinfo{author}{\bibfnamefont{P.}~\bibnamefont{H\"{a}nggi}},
  \bibinfo{journal}{Phys. Rev. E} \textbf{\bibinfo{volume}{61}},
  \bibinfo{pages}{R4687} (\bibinfo{year}{2000}).

\bibitem[{\citenamefont{Falci et~al.}(2005)\citenamefont{Falci, D'Arrigo,
  Mastellone, and Paladino}}]{Falci05}
\bibinfo{author}{\bibfnamefont{G.}~\bibnamefont{Falci}},
  \bibinfo{author}{\bibfnamefont{A.}~\bibnamefont{D'Arrigo}},
  \bibinfo{author}{\bibfnamefont{A.}~\bibnamefont{Mastellone}},
  \bibnamefont{and} \bibinfo{author}{\bibfnamefont{E.}~\bibnamefont{Paladino}},
  \bibinfo{journal}{Phys. Rev. Lett.} \textbf{\bibinfo{volume}{94}},
  \bibinfo{pages}{167002} (\bibinfo{year}{2005}).

\bibitem[{\citenamefont{Nakamura et~al.}(2002)\citenamefont{Nakamura, {Yu. A.
  Pashkin}, Yamamoto, and Tsai}}]{Nakamura02}
\bibinfo{author}{\bibfnamefont{Y.}~\bibnamefont{Nakamura}},
  \bibinfo{author}{\bibnamefont{{Yu. A. Pashkin}}},
  \bibinfo{author}{\bibfnamefont{T.}~\bibnamefont{Yamamoto}}, \bibnamefont{and}
  \bibinfo{author}{\bibfnamefont{J.~S.} \bibnamefont{Tsai}},
  \bibinfo{journal}{Phys. Rev. Lett.} \textbf{\bibinfo{volume}{88}},
  \bibinfo{pages}{047901} (\bibinfo{year}{2002}).

\bibitem[{\citenamefont{Lehnert et~al.}(2003)\citenamefont{Lehnert, Bladh,
  Spietz, Gunnarson, Schuster, Delsing, and Schoelkopf}}]{Lehnert03}
\bibinfo{author}{\bibfnamefont{K.~W.} \bibnamefont{Lehnert}},
  \bibinfo{author}{\bibfnamefont{K.}~\bibnamefont{Bladh}},
  \bibinfo{author}{\bibfnamefont{L.~F.} \bibnamefont{Spietz}},
  \bibinfo{author}{\bibfnamefont{D.}~\bibnamefont{Gunnarson}},
  \bibinfo{author}{\bibfnamefont{D.~I.} \bibnamefont{Schuster}},
  \bibinfo{author}{\bibfnamefont{P.}~\bibnamefont{Delsing}}, \bibnamefont{and}
  \bibinfo{author}{\bibfnamefont{R.~J.} \bibnamefont{Schoelkopf}},
  \bibinfo{journal}{Phys. Rev. Lett.} \textbf{\bibinfo{volume}{90}},
  \bibinfo{pages}{027002} (\bibinfo{year}{2003}).

\bibitem[{\citenamefont{Duty et~al.}(2004)\citenamefont{Duty, Gunnarsson,
  Bladh, and Delsing}}]{Duty04}
\bibinfo{author}{\bibfnamefont{T.}~\bibnamefont{Duty}},
  \bibinfo{author}{\bibfnamefont{D.}~\bibnamefont{Gunnarsson}},
  \bibinfo{author}{\bibfnamefont{K.}~\bibnamefont{Bladh}}, \bibnamefont{and}
  \bibinfo{author}{\bibfnamefont{P.}~\bibnamefont{Delsing}},
  \bibinfo{journal}{Phys. Rev. B} \textbf{\bibinfo{volume}{69}},
  \bibinfo{pages}{140503(R)} (\bibinfo{year}{2004}).

\bibitem[{\citenamefont{Li et~al.}(2005)\citenamefont{Li, Qiu, Zhou, Matheny,
  Chen, Lukens, and Han}}]{Lishaoxiong06}
\bibinfo{author}{\bibfnamefont{S.-X.} \bibnamefont{Li}},
  \bibinfo{author}{\bibfnamefont{W.}~\bibnamefont{Qiu}},
  \bibinfo{author}{\bibfnamefont{Z.}~\bibnamefont{Zhou}},
  \bibinfo{author}{\bibfnamefont{M.}~\bibnamefont{Matheny}},
  \bibinfo{author}{\bibfnamefont{W.}~\bibnamefont{Chen}},
  \bibinfo{author}{\bibfnamefont{J.~E.} \bibnamefont{Lukens}},
  \bibnamefont{and} \bibinfo{author}{\bibfnamefont{S.}~\bibnamefont{Han}},
  \bibinfo{journal}{arXiv: cond-mat/0507008}  (\bibinfo{year}{2005}).

\bibitem[{\citenamefont{Bertet et~al.}(unpublished)\citenamefont{Bertet,
  Chiorescu, Burkard, Semba, Harmans, DiVincenzo, and Mooij}}]{Bertet06}
\bibinfo{author}{\bibfnamefont{P.}~\bibnamefont{Bertet}},
  \bibinfo{author}{\bibfnamefont{I.}~\bibnamefont{Chiorescu}},
  \bibinfo{author}{\bibfnamefont{G.}~\bibnamefont{Burkard}},
  \bibinfo{author}{\bibfnamefont{K.}~\bibnamefont{Semba}},
  \bibinfo{author}{\bibfnamefont{C.~J. P.~M.} \bibnamefont{Harmans}},
  \bibinfo{author}{\bibfnamefont{D.}~\bibnamefont{DiVincenzo}},
  \bibnamefont{and} \bibinfo{author}{\bibfnamefont{J.~E.} \bibnamefont{Mooij}},
  \bibinfo{journal}{arXiv:cond-mat/0412485}  (\bibinfo{year}{unpublished}).

\bibitem[{\citenamefont{Berkley
  et~al.}(2003{\natexlab{b}})\citenamefont{Berkley, Xu, Gubrud, Ramos,
  Anderson, Lobb, and Wellstood}}]{Berkley03-1}
\bibinfo{author}{\bibfnamefont{A.~J.} \bibnamefont{Berkley}},
  \bibinfo{author}{\bibfnamefont{H.}~\bibnamefont{Xu}},
  \bibinfo{author}{\bibfnamefont{M.~A.} \bibnamefont{Gubrud}},
  \bibinfo{author}{\bibfnamefont{R.~C.} \bibnamefont{Ramos}},
  \bibinfo{author}{\bibfnamefont{J.~R.} \bibnamefont{Anderson}},
  \bibinfo{author}{\bibfnamefont{C.~J.} \bibnamefont{Lobb}}, \bibnamefont{and}
  \bibinfo{author}{\bibfnamefont{F.~C.} \bibnamefont{Wellstood}},
  \bibinfo{journal}{Phys. Rev. B} \textbf{\bibinfo{volume}{68}},
  \bibinfo{pages}{060502(R)} (\bibinfo{year}{2003}{\natexlab{b}}).

\bibitem[{\citenamefont{Dutta et~al.}(2004)\citenamefont{Dutta, Xu, Berkley,
  Ramos, Gubrud, Anderson, Lobb, and Wellstood}}]{Dutta2004}
\bibinfo{author}{\bibfnamefont{S.~K.} \bibnamefont{Dutta}},
  \bibinfo{author}{\bibfnamefont{H.}~\bibnamefont{Xu}},
  \bibinfo{author}{\bibfnamefont{A.~J.} \bibnamefont{Berkley}},
  \bibinfo{author}{\bibfnamefont{R.~C.} \bibnamefont{Ramos}},
  \bibinfo{author}{\bibfnamefont{M.~A.} \bibnamefont{Gubrud}},
  \bibinfo{author}{\bibfnamefont{J.~R.} \bibnamefont{Anderson}},
  \bibinfo{author}{\bibfnamefont{C.~J.} \bibnamefont{Lobb}}, \bibnamefont{and}
  \bibinfo{author}{\bibfnamefont{F.~C.} \bibnamefont{Wellstood}},
  \bibinfo{journal}{Phys. Rev. B} \textbf{\bibinfo{volume}{70}},
  \bibinfo{pages}{140502(R)} (\bibinfo{year}{2004}).

\bibitem[{\citenamefont{Viola and Knill}(2005)}]{Viola05}
\bibinfo{author}{\bibfnamefont{L.}~\bibnamefont{Viola}} \bibnamefont{and}
  \bibinfo{author}{\bibfnamefont{E.}~\bibnamefont{Knill}},
  \bibinfo{journal}{Phys. Rev. Lett.} \textbf{\bibinfo{volume}{94}},
  \bibinfo{pages}{060502} (\bibinfo{year}{2005}).

\bibitem[{\citenamefont{Viola and Lloyd}(1998)}]{Viola98}
\bibinfo{author}{\bibfnamefont{L.}~\bibnamefont{Viola}} \bibnamefont{and}
  \bibinfo{author}{\bibfnamefont{S.}~\bibnamefont{Lloyd}},
  \bibinfo{journal}{Phys. Rev. A} \textbf{\bibinfo{volume}{58}},
  \bibinfo{pages}{2733} (\bibinfo{year}{1998}).

\bibitem[{\citenamefont{Viola et~al.}(1999)\citenamefont{Viola, Knill, and
  Lloyd}}]{Viola99}
\bibinfo{author}{\bibfnamefont{L.}~\bibnamefont{Viola}},
  \bibinfo{author}{\bibfnamefont{E.}~\bibnamefont{Knill}}, \bibnamefont{and}
  \bibinfo{author}{\bibfnamefont{S.}~\bibnamefont{Lloyd}},
  \bibinfo{journal}{Phys. Rev. Lett.} \textbf{\bibinfo{volume}{82}},
  \bibinfo{pages}{2417} (\bibinfo{year}{1999}).

\bibitem[{\citenamefont{Vitali and Tombesi}(2002)}]{Vitali02}
\bibinfo{author}{\bibfnamefont{D.}~\bibnamefont{Vitali}} \bibnamefont{and}
  \bibinfo{author}{\bibfnamefont{P.}~\bibnamefont{Tombesi}},
  \bibinfo{journal}{Phys. Rev. A} \textbf{\bibinfo{volume}{65}},
  \bibinfo{pages}{012305} (\bibinfo{year}{2002}).

\bibitem[{\citenamefont{Gutmann et~al.}(2005)\citenamefont{Gutmann, Wilhelm,
  Kaminsky, and Lloyd}}]{Gutmann05}
\bibinfo{author}{\bibfnamefont{H.}~\bibnamefont{Gutmann}},
  \bibinfo{author}{\bibfnamefont{F.~K.} \bibnamefont{Wilhelm}},
  \bibinfo{author}{\bibfnamefont{W.~M.} \bibnamefont{Kaminsky}},
  \bibnamefont{and} \bibinfo{author}{\bibfnamefont{S.}~\bibnamefont{Lloyd}},
  \bibinfo{journal}{Phys. Rev. A} \textbf{\bibinfo{volume}{71}},
  \bibinfo{pages}{020302(R)} (\bibinfo{year}{2005}).

\bibitem[{\citenamefont{Falci et~al.}(2004)\citenamefont{Falci, D'Arrigo,
  Mastellone, and Paladino}}]{Falci04}
\bibinfo{author}{\bibfnamefont{G.}~\bibnamefont{Falci}},
  \bibinfo{author}{\bibfnamefont{A.}~\bibnamefont{D'Arrigo}},
  \bibinfo{author}{\bibfnamefont{A.}~\bibnamefont{Mastellone}},
  \bibnamefont{and} \bibinfo{author}{\bibfnamefont{E.}~\bibnamefont{Paladino}},
  \bibinfo{journal}{Phys. Rev. A} \textbf{\bibinfo{volume}{70}},
  \bibinfo{pages}{040101(R)} (\bibinfo{year}{2004}).

\bibitem[{\citenamefont{Faoro and Viola}(2004)}]{Faoro04}
\bibinfo{author}{\bibfnamefont{L.}~\bibnamefont{Faoro}} \bibnamefont{and}
  \bibinfo{author}{\bibfnamefont{L.}~\bibnamefont{Viola}},
  \bibinfo{journal}{Phys. Rev. Lett.} \textbf{\bibinfo{volume}{92}},
  \bibinfo{pages}{117905} (\bibinfo{year}{2004}).

\bibitem[{\citenamefont{Shiokawa and Lidar}(2004)}]{Shiokawa04}
\bibinfo{author}{\bibfnamefont{K.}~\bibnamefont{Shiokawa}} \bibnamefont{and}
  \bibinfo{author}{\bibfnamefont{D.~A.} \bibnamefont{Lidar}},
  \bibinfo{journal}{Phys. Rev. A} \textbf{\bibinfo{volume}{69}},
  \bibinfo{pages}{030302(R)} (\bibinfo{year}{2004}).

\bibitem[{\citenamefont{Duan and Guo}(1997)}]{Duan97}
\bibinfo{author}{\bibfnamefont{L.-M.} \bibnamefont{Duan}} \bibnamefont{and}
  \bibinfo{author}{\bibfnamefont{G.-C.} \bibnamefont{Guo}},
  \bibinfo{journal}{Phys. Rev. Lett.} \textbf{\bibinfo{volume}{79}},
  \bibinfo{pages}{1953} (\bibinfo{year}{1997}).

\bibitem[{\citenamefont{Zanardi and Rasetti}(1997)}]{Zanardi97}
\bibinfo{author}{\bibfnamefont{P.}~\bibnamefont{Zanardi}} \bibnamefont{and}
  \bibinfo{author}{\bibfnamefont{M.}~\bibnamefont{Rasetti}},
  \bibinfo{journal}{Phys. Rev. Lett.} \textbf{\bibinfo{volume}{79}},
  \bibinfo{pages}{3306} (\bibinfo{year}{1997}).

\bibitem[{\citenamefont{Lidar et~al.}(1998)\citenamefont{Lidar, Chuang, and
  Whaley}}]{Lidar98}
\bibinfo{author}{\bibfnamefont{D.~A.} \bibnamefont{Lidar}},
  \bibinfo{author}{\bibfnamefont{I.~L.} \bibnamefont{Chuang}},
  \bibnamefont{and} \bibinfo{author}{\bibfnamefont{K.~B.}
  \bibnamefont{Whaley}}, \bibinfo{journal}{Phys. Rev. Lett.}
  \textbf{\bibinfo{volume}{81}}, \bibinfo{pages}{2594} (\bibinfo{year}{1998}).

\bibitem[{\citenamefont{Beige et~al.}(2000)\citenamefont{Beige, Braun,
  Tregenna, and Knight}}]{Beige00}
\bibinfo{author}{\bibfnamefont{A.}~\bibnamefont{Beige}},
  \bibinfo{author}{\bibfnamefont{D.}~\bibnamefont{Braun}},
  \bibinfo{author}{\bibfnamefont{B.}~\bibnamefont{Tregenna}}, \bibnamefont{and}
  \bibinfo{author}{\bibfnamefont{P.~L.} \bibnamefont{Knight}},
  \bibinfo{journal}{Phys. Rev. Lett.} \textbf{\bibinfo{volume}{85}},
  \bibinfo{pages}{1762} (\bibinfo{year}{2000}).

\bibitem[{\citenamefont{Bacon et~al.}(2001)\citenamefont{Bacon, Brown, and
  Whaley}}]{Bacon01}
\bibinfo{author}{\bibfnamefont{D.}~\bibnamefont{Bacon}},
  \bibinfo{author}{\bibfnamefont{K.~R.} \bibnamefont{Brown}}, \bibnamefont{and}
  \bibinfo{author}{\bibfnamefont{K.~B.} \bibnamefont{Whaley}},
  \bibinfo{journal}{Phys. Rev. Lett.} \textbf{\bibinfo{volume}{87}},
  \bibinfo{pages}{247902} (\bibinfo{year}{2001}).

\bibitem[{\citenamefont{Zhou et~al.}(2005)\citenamefont{Zhou, Chu, and
  Han}}]{zhou2005PRL-ITA}
\bibinfo{author}{\bibfnamefont{Z.}~\bibnamefont{Zhou}},
  \bibinfo{author}{\bibfnamefont{S.-I.} \bibnamefont{Chu}}, \bibnamefont{and}
  \bibinfo{author}{\bibfnamefont{S.}~\bibnamefont{Han}},
  \bibinfo{journal}{Phys. Rev. Lett.} \textbf{\bibinfo{volume}{95}},
  \bibinfo{pages}{120501} (\bibinfo{year}{2005}).

\bibitem[{\citenamefont{Zhou et~al.}(2006)\citenamefont{Zhou, Chu, and
  Han}}]{zhou2006}
\bibinfo{author}{\bibfnamefont{Z.}~\bibnamefont{Zhou}},
  \bibinfo{author}{\bibfnamefont{S.-I.} \bibnamefont{Chu}}, \bibnamefont{and}
  \bibinfo{author}{\bibfnamefont{S.}~\bibnamefont{Han}},
  \bibinfo{journal}{Phys. Rev. B} \textbf{\bibinfo{volume}{73}},
  \bibinfo{pages}{104521} (\bibinfo{year}{2006}).

\bibitem[{\citenamefont{Burkard et~al.}(2004)\citenamefont{Burkard, Koch, and
  DiVincenzo}}]{Burkard2004}
\bibinfo{author}{\bibfnamefont{G.}~\bibnamefont{Burkard}},
  \bibinfo{author}{\bibfnamefont{R.~H.} \bibnamefont{Koch}}, \bibnamefont{and}
  \bibinfo{author}{\bibfnamefont{D.~P.} \bibnamefont{DiVincenzo}},
  \bibinfo{journal}{Phys. Rev. B} \textbf{\bibinfo{volume}{69}},
  \bibinfo{pages}{064503} (\bibinfo{year}{2004}).

\bibitem[{\citenamefont{Brinati et~al.}(1994)\citenamefont{Brinati, Mizrahi,
  and Prataviera}}]{Brinati94}
\bibinfo{author}{\bibfnamefont{J.~R.} \bibnamefont{Brinati}},
  \bibinfo{author}{\bibfnamefont{S.~S.} \bibnamefont{Mizrahi}},
  \bibnamefont{and} \bibinfo{author}{\bibfnamefont{G.~A.}
  \bibnamefont{Prataviera}}, \bibinfo{journal}{Phys. Rev. A}
  \textbf{\bibinfo{volume}{50}}, \bibinfo{pages}{3304} (\bibinfo{year}{1994}).

\bibitem[{\citenamefont{{Yu. Smirnov}}(2003)}]{Smirnov2003}
\bibinfo{author}{\bibfnamefont{A.}~\bibnamefont{{Yu. Smirnov}}},
  \bibinfo{journal}{Phys. Rev. B} \textbf{\bibinfo{volume}{67}},
  \bibinfo{pages}{155104} (\bibinfo{year}{2003}).

\bibitem[{\citenamefont{Louisell}(1973)}]{Louisell1973}
\bibinfo{author}{\bibfnamefont{W.~H.} \bibnamefont{Louisell}},
  \emph{\bibinfo{title}{Quantum Statistical Properties of Radiation}}
  (\bibinfo{publisher}{John Wiley \&\ Sons}, \bibinfo{year}{1973}).

\bibitem[{\citenamefont{Weiss}(1999)}]{Weiss1999}
\bibinfo{author}{\bibfnamefont{U.}~\bibnamefont{Weiss}},
  \emph{\bibinfo{title}{Quantum Dissipative Systems}}
  (\bibinfo{publisher}{World Scientific Publishing Co., Singapore},
  \bibinfo{year}{1999}), \bibinfo{edition}{2nd} ed.

\bibitem[{\citenamefont{Zhou et~al.}(to be submitted)\citenamefont{Zhou, Chu,
  and Han}}]{Zhou-PRB-I}
\bibinfo{author}{\bibfnamefont{Z.}~\bibnamefont{Zhou}},
  \bibinfo{author}{\bibfnamefont{S.-I.} \bibnamefont{Chu}}, \bibnamefont{and}
  \bibinfo{author}{\bibfnamefont{S.}~\bibnamefont{Han}},
  \bibinfo{journal}{Phys. Rev. B}  (\bibinfo{year}{to be submitted}).

\bibitem[{\citenamefont{Leggett et~al.}(1987)\citenamefont{Leggett,
  Chakravarty, Dorsey, Fisher, Garg, and Zwerger}}]{Leggett87}
\bibinfo{author}{\bibfnamefont{A.}~\bibnamefont{Leggett}},
  \bibinfo{author}{\bibfnamefont{S.}~\bibnamefont{Chakravarty}},
  \bibinfo{author}{\bibfnamefont{A.}~\bibnamefont{Dorsey}},
  \bibinfo{author}{\bibfnamefont{M.}~\bibnamefont{Fisher}},
  \bibinfo{author}{\bibfnamefont{A.}~\bibnamefont{Garg}}, \bibnamefont{and}
  \bibinfo{author}{\bibfnamefont{W.}~\bibnamefont{Zwerger}},
  \bibinfo{journal}{Rev. Mod. Phys.} \textbf{\bibinfo{volume}{59}},
  \bibinfo{pages}{1} (\bibinfo{year}{1987}).

\bibitem[{\citenamefont{Devoret}(1997)}]{Devoret97}
\bibinfo{author}{\bibfnamefont{M.~H.} \bibnamefont{Devoret}}, in
  \emph{\bibinfo{booktitle}{Quantum Fluctuations}}, edited by
  \bibinfo{editor}{\bibfnamefont{S.}~\bibnamefont{Reynaud}},
  \bibinfo{editor}{\bibfnamefont{E.}~\bibnamefont{Giacobino}},
  \bibnamefont{and}
  \bibinfo{editor}{\bibfnamefont{J.}~\bibnamefont{Zinn-Justin}},
  \bibinfo{organization}{Les Houches, France, 27 June-28 July 1995}
  (\bibinfo{publisher}{Elsevier Science B. V.}, \bibinfo{year}{1997}), pp.
  \bibinfo{pages}{351--386}.

\bibitem[{\citenamefont{Grifoni et~al.}(1999)\citenamefont{Grifoni, Paladino,
  and U.Weiss}}]{Grifoni1999}
\bibinfo{author}{\bibfnamefont{M.}~\bibnamefont{Grifoni}},
  \bibinfo{author}{\bibfnamefont{E.}~\bibnamefont{Paladino}}, \bibnamefont{and}
  \bibinfo{author}{\bibnamefont{U.Weiss}}, \bibinfo{journal}{Eur. J. Phys. B}
  \textbf{\bibinfo{volume}{10}}, \bibinfo{pages}{719} (\bibinfo{year}{1999}).

\bibitem[{\citenamefont{Kosugi et~al.}(2005)\citenamefont{Kosugi, Matsuo,
  Konno, and Hatakenaka}}]{Kosugi05}
\bibinfo{author}{\bibfnamefont{N.}~\bibnamefont{Kosugi}},
  \bibinfo{author}{\bibfnamefont{S.}~\bibnamefont{Matsuo}},
  \bibinfo{author}{\bibfnamefont{K.}~\bibnamefont{Konno}}, \bibnamefont{and}
  \bibinfo{author}{\bibfnamefont{N.}~\bibnamefont{Hatakenaka}},
  \bibinfo{journal}{Phys. Rev. B} \textbf{\bibinfo{volume}{72}},
  \bibinfo{pages}{172509} (\bibinfo{year}{2005}).

\bibitem[{\citenamefont{Danilov et~al.}(1983)\citenamefont{Danilov, Likharev,
  and Zorin}}]{Danilov1983}
\bibinfo{author}{\bibfnamefont{V.~V.} \bibnamefont{Danilov}},
  \bibinfo{author}{\bibfnamefont{K.}~\bibnamefont{Likharev}}, \bibnamefont{and}
  \bibinfo{author}{\bibfnamefont{A.~B.} \bibnamefont{Zorin}},
  \bibinfo{journal}{IEEE Trans. Magn.} \textbf{\bibinfo{volume}{19}},
  \bibinfo{pages}{572} (\bibinfo{year}{1983}).

\bibitem[{\citenamefont{Zhou et~al.}(2002)\citenamefont{Zhou, Chu, and
  Han}}]{Zhou2002}
\bibinfo{author}{\bibfnamefont{Z.}~\bibnamefont{Zhou}},
  \bibinfo{author}{\bibfnamefont{S.-I.} \bibnamefont{Chu}}, \bibnamefont{and}
  \bibinfo{author}{\bibfnamefont{S.}~\bibnamefont{Han}},
  \bibinfo{journal}{Phys. Rev. B} \textbf{\bibinfo{volume}{66}},
  \bibinfo{pages}{054527} (\bibinfo{year}{2002}).

\bibitem[{\citenamefont{Hermann and {Fleck, Jr.}}(1988)}]{Hermann1988}
\bibinfo{author}{\bibfnamefont{M.~R.} \bibnamefont{Hermann}} \bibnamefont{and}
  \bibinfo{author}{\bibfnamefont{J.~A.} \bibnamefont{{Fleck, Jr.}}},
  \bibinfo{journal}{Phys. Rev. A} \textbf{\bibinfo{volume}{38}},
  \bibinfo{pages}{6000} (\bibinfo{year}{1988}).

\bibitem[{\citenamefont{Alekseev and Sushilov}(1992)}]{Alekseev1992}
\bibinfo{author}{\bibfnamefont{A.~V.} \bibnamefont{Alekseev}} \bibnamefont{and}
  \bibinfo{author}{\bibfnamefont{N.~V.} \bibnamefont{Sushilov}},
  \bibinfo{journal}{Phys. Rev. A} \textbf{\bibinfo{volume}{46}},
  \bibinfo{pages}{351} (\bibinfo{year}{1992}).

\end{thebibliography}

\bigskip \newpage

\textbf{Figure Captions}

FIG. 1 (a) A simplified external circuit of the SQUID flux qubit. The left
part is the SQUID qubit and the right part is the circuit used to supply
external flux to the SQUID qubit. $I_{1}$ and $V_{1}$ are the current and
voltage of the SQUID qubit's circuit, $I_{2}$ is the current of the external
circuit, $L_{e}$ is the inductance of the superconducting loop of the
external circuit, $M$ is the mutual inductance of the SQUID qubit and
external circuit, and $Y_{e}(\omega )$ is the effective admittance of other
devices in the external circuit. (b) The equivalent admittance $Y\left(
\omega \right) $ of the external circuit.

\bigskip

FIG. 2 (Online color) $J(\omega )$ vs. $\omega $\ for the SQUID flux qubit
at $T=0.1$ K.

\bigskip

FIG. 3 (Online color) $J(\omega )$ vs. $T$ for the SQUID flux qubit at $%
\omega =\omega _{LC}$. The dashed and dashed dotted lines are the spectral
densities at low temperature limit ($T\rightarrow 0$) and high temperature
limit ($T\rightarrow \infty $), respectively.

\bigskip

FIG. 4 (Online color) Evolution of (a) the population inversion and (b) the
absolutely squared coherence of the SQUID flux qubit in free decay. The
solid and dashed lines are the numerical and fitting results, respectively.

\bigskip

FIG. 5 (Online color) Evolution of (a) the population difference and (b) the
absolutely squared coherence of the microwave-driven SQUID flux qubit. The
solid and dashed lines are the numerical and fitting results, respectively.

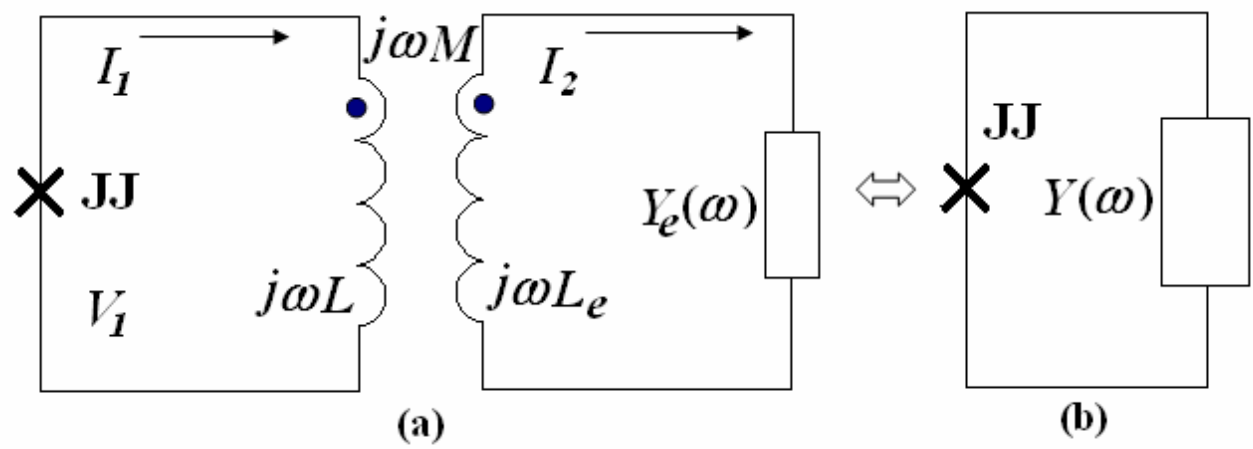
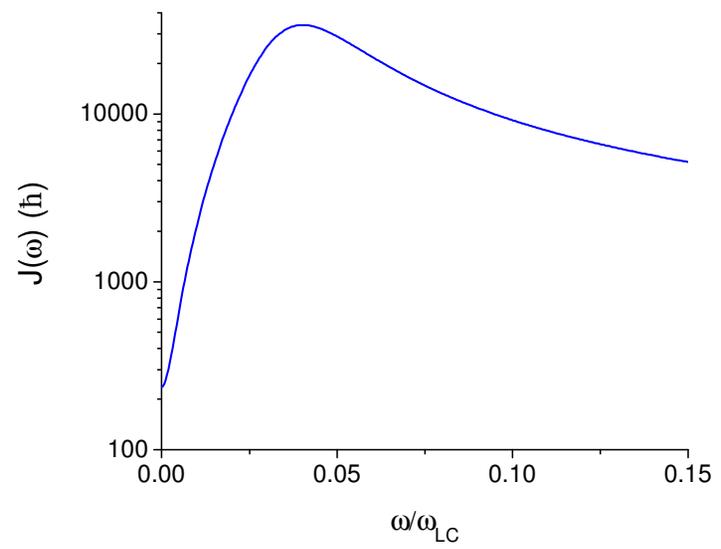
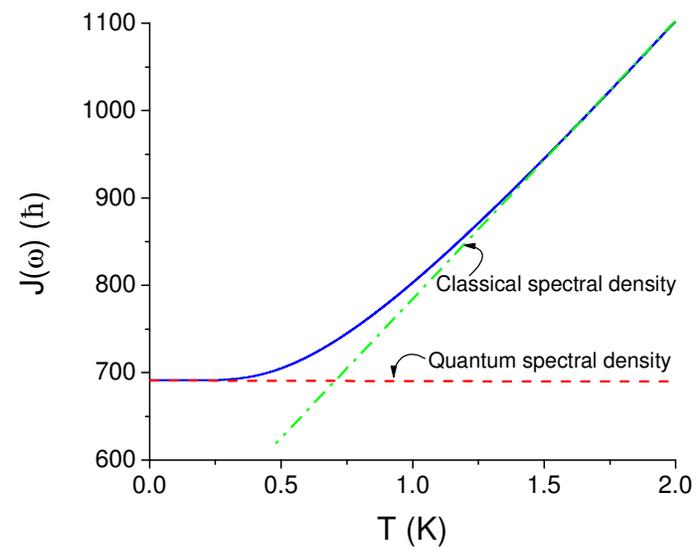
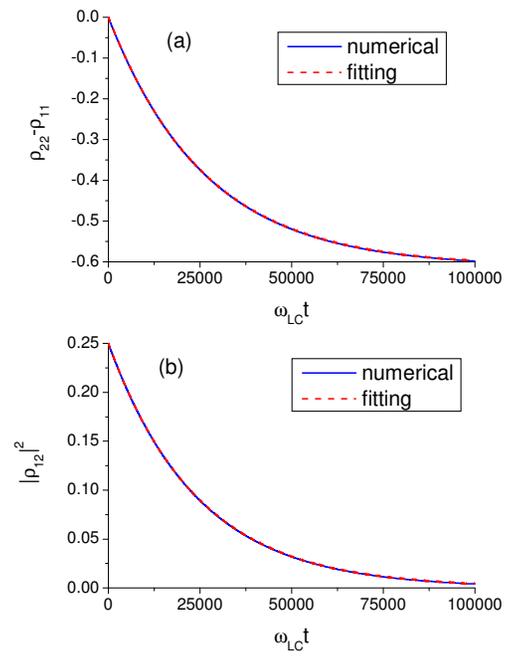
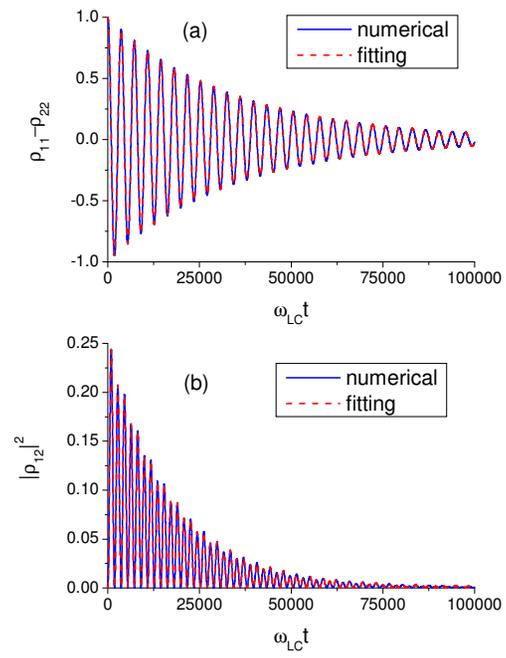
\end{document}